\documentclass[fleqn,10pt]{wlscirep}
\usepackage[utf8]{inputenc}
\usepackage[T1]{fontenc}
\usepackage{lineno}
\usepackage{multirow}
\usepackage{amsmath}
\usepackage{siunitx}
\usepackage{tikz}
\usepackage{makecell}
\usepackage{xurl}

\usepackage{graphicx}
\usepackage{xcolor}
\usepackage{caption}
\usepackage{subcaption}
\usepackage{comment}
\usepackage{tabularx}

\usepackage{colortbl}
\usepackage{adjustbox}
\usepackage{array}
\usepackage{booktabs} 
\usepackage{nicematrix}
 \usepackage{xfrac}

\usepackage{siunitx}

\title{UrBAN: Urban Beehive Acoustics and PheNotyping Dataset} 

\author[1,*]{Mahsa Abdollahi}
\author[1]{Yi Zhu}
\author[1]{Heitor R. Guimarães}

\author[3]{Nico Coallier}
\author[2]{Ségolène Maucourt}
\author[2]{Pierre Giovenazzo}
\author[1]{Tiago H. Falk}

\affil[1]{INRS-EMT, Université du Québec, Montréal, Canada}
\affil[2]{Département de biologie, Université Laval, Laval, Canada}
\affil[3]{Nectar Technologies Inc., Montréal, Canada}

\affil[*]{corresponding author(s): tiago.falk@inrs.ca}

\begin{abstract}
 
In this paper, we present a multimodal dataset obtained from a honey bee colony in Montréal, Quebec, Canada, spanning the years of 2021 to 2022. This apiary comprised 10 beehives, with microphones recording more than 2000 hours of high quality raw audio, and also sensors capturing temperature, and humidity. Periodic hive inspections involved monitoring colony honey bee population changes, assessing queen-related conditions, and documenting overall hive health. Additionally, health metrics, such as Varroa mite infestation rates and winter mortality assessments were recorded, offering valuable insights into factors affecting hive health status and resilience. In this study, we first outline the data collection process, sensor data description, and dataset structure. Furthermore, we demonstrate a practical application of this dataset by extracting various features from the raw audio to predict colony population using the number of frames of bees as a proxy.

\end{abstract}

\begin{document}

\flushbottom
\maketitle
\thispagestyle{empty}

\section*{Background and Summary} 
Honey bees ({\it{Apis mellifera}}) play a critical role in ecological balance, serving as essential pollinators for both agricultural crops and natural biodiversity. Their contributions extend beyond honey and beeswax production, impacting many fruit and seed crops, including almonds, citrus fruits, and blueberries. This dependence underscores the significance of honey bee populations for food production sustainability and quality. However, declines in bee colony health and population size can have far-reaching consequences for the agricultural industry. 

Conventional methods of beehive surveillance rely on manual and visual inspections, which are labor-intensive for beekeepers and disruptive to colonies, resulting in infrequent checks. Typically, beekeepers examine their hives every two weeks during active periods of pollination or honey production, and less frequently during winter. However, significant changes in colony dynamics and health can occur within this time frame, thus necessitating continuous monitoring. Global reports of substantial colony losses underscore the urgency of this issue. 

In recent years, there has been a global observation of significant colony losses, which have been attributed to various stressors working either independently or in combination. These stressors include pesticides, pathogens, parasites, climate variations, as well as other factors~\cite{french2024honey, gaubert2023individual, brodschneider2018multi, gray2019loss, gray2020honey}. Consequently, passive monitoring of honeybee colony health has attracted significant attention from the beekeeping and the research communities.

Recently, with the advancement of IoT (Internet-of-Things) in precision beekeeping, automated beehive monitoring tools have emerged to overcome the shortcomings of human manual inspections and colonies management~\cite{cota2023bhivesense}. These systems typically deploy sensors within the hive to monitor real-time colony status and assess its condition~\cite{abdollahi2022automated, pb_2, alleri2023recent}. Existing systems typically gather information such as temperature, humidity, beehive weight, and acoustics. For example, temperature stability within the hive is crucial for bee health and brood development, thus directly impacting hive productivity~\cite{cetin2004effects, seeley1981regulation, seeley2014honeybee}. Moreover, relative humidity affects larval growth, colony development, and bee behavior, with variations influencing water transportation and feeding~\cite{abou2017review, human2006honeybees}. Moreover, hive weight is an essential measurement for researchers, offering insights into colony activities, such as nectar collection and food consumption, showing variations over the course of the day. Continuous monitoring of colony weight, particularly a reference colony, aids in identifying the start and end of nectar flow and in assessing colony foraging activity~\cite{meikle2008within, zacepins2020monitoring}. 

While temperature and humidity can provide some complimentary information about the health of a colony, beehive acoustics have proven to be a more effective method, as bees communicate internally using vibrations and acoustic signals, generated through body movements, wing flapping, and muscle contractions~\cite{hunt2013intracolony}. These signals include sounds associated with different events, such as mite attacks, queen failure, and swarming, making them an ideal modality for beehive monitoring. In the literature, various acoustic monitoring systems have been highlighted, offering capabilities such as queen absence detection~\cite{rustam2024bee, ieee_1, ieee_16, ieee_26}, bee activity~\cite{ieee_2, arxiv_3}, swarming\cite{ieee_7, ieee_18, sc_7}, hive strength~\cite{arxiv_1}, pathogen/parasite infestations~\cite{sc_34}, environmental pollutants~\cite{sc_2, ref_2}, and early prediction of colony winter survivability~\cite{zhu2023bee}. 

Open access data on beehive management is crucial for fostering research output and advancements in the field, especially with the recent advances seen in artificial intelligence (AI) and machine learning (ML). Unfortunately, the beekeeping community faces a significant scarcity of open access data, hindering the progress of research and innovation. Access to comprehensive datasets on beehive dynamics, including factors such as colony health, behavior, and environmental influences, empowers researchers to conduct in-depth analyses and develop impactful solutions. While there are some public bee audio database available such as Buzz~\cite{kulyukin2021audio}, Nu-hive~\cite{cecchi2018preliminary}, and OSBH~\cite{osbh,arxiv_3}, they are mostly focused on bee and queen bee detection and are limited in sample size. To bee or not to bee dataset~\cite{arxiv_3} is a combination of Nu-hive and osbh audio samples, labeled for a different purpose which is detecting bee buzzing sound. 

Recently, the MSPB dataset was released~\cite{zhu2023mspb}; this is a longitudinal multi-sensor dataset with phenotypic measurements has been released showing that how audio signals can be used for detection of winter survivability and population estimation. The dataset, however, does not provide access to the raw audio signals and makes available only pre-processed parameters extracted from the audio signal (e.g., overall hive audio power). Table~\ref{tab:public_datasets} lists available beehive acoustic datasets along with some relevant statistics. It can be seen from Table~\ref{tab:public_datasets} that the existing datasets are limited in size of raw audio samples and the days that they covered. Therefore, to address these limitations and support advancements in beehive monitoring, we present a new beehive acoustics dataset.

\begin{table}[]
    \centering
    \begin{NiceTabular}{c|c|c|c|c|c}
        \toprule
        Name &  \# Hives & Labels & Modality & Size  & Availability\\
        \midrule
        \multirow{3}{*}{NU-Hive~\cite{cecchi2018preliminary}}  &  \multirow{3}{*}{2} & \multirow{3}{*}{Queenright/queenless} & Raw audio &  \SI{96}{h} & Public\\
        \cmidrule{4-6}
         &   &  & Temperature & - & Not available\\
        \cmidrule{4-6}
          &   & &Humidity & - & Not available\\
        \midrule
        BUZZ~\cite{kulyukin2021audio} & 6 & Bee buzzing/cricket/noise &Raw audio & \SI{7}{h} & Upon requests\\
        \midrule
        OSBH~\cite{osbh, arxiv_3} & 6 & Queenright/queenless & Raw audio & 140 min & Subset available\\
        \midrule
        \makecell{Too bee\\ or\\ not to bee~\cite{arxiv_3}} & 8 & Bee buzzing/no bee buzzing & Raw audio & \SI{12}{h}& Public\\
        \midrule
        \multirow{3}{*}{MSPB~\cite{zhu2023mspb}}  &  \multirow{3}{*}{53} & \multirow{3}{*}{\shortstack{Population; Honey yield; \\ Queen conditions;\\ Hygienic behavior;\\ Winter mortality}} & \makecell{Hand crafted \\audio features} &\multirow{3}{*}{\makecell{Quarter-hourly  \\ during 365 day}} & \multirow{3}{*}{Public}\\
        \cmidrule{4-4}
         &   &  & Temperature &  & \\
        \cmidrule{4-4}
          &   & &Humidity &  & \\
        \midrule
        \multirow{4}{*}{\makecell{Smart Bee\\ Colony Monitor~\cite{smart_bee}}} & \multirow{4}{*}{4} & \multirow{4}{*}{Queenright/queenless} & Raw audio & \SI{118}{h} & \multirow{4}{*}{Public}\\
        \cmidrule{4-5}
         &   &  & Temperature & \multirow{3}{*}{\makecell{Hourly during \\ 40 day}} & \\
        \cmidrule{4-4}
          &   & & Humidity &  & \\
          \cmidrule{4-4}
          &   & & Pressure &  & \\
        \midrule
        \multirow{3}{*}{UrBAN (ours)}  &  \multirow{3}{*}{10} & \multirow{3}{*}{\shortstack{Population; \\ Queen conditions;\\Winter mortality}}& Raw audio &  \SI{3171}{h} &\multirow{3}{*}{Public}\\
        \cmidrule{4-5}
         &   &  & Temperature & \multirow{2}{*}{\makecell{Quarter-hourly \\during 135 day}} & \\
        \cmidrule{4-4}
          &   & &Humidity &  & \\
        \bottomrule
    \end{NiceTabular}
    \caption{Comparison of UrBAN dataset with the other public datasets.}
    \label{tab:public_datasets}
\end{table}

In this paper, we introduce the UrBAN (Urban Beehive Acoustics and PheNotyping) dataset that includes over 2000 hours of raw audio samples collected from beehives during a period of two years. The main focus of the data was on colony population prediction. The population of a colony can be estimated using the number of frames of bees covered by least 70\%~\cite{chabert2021rapid}. There have been some studies showing that the number of frames of bees can be predicted using various features extracted from the raw audio within a machine learning framework~\cite{abdollahi2022importance, zhu2023mspb, di2023applicability}. Authors in~\cite{zhu2023mspb} showed that features such as audio power in specific band of frequency and its variation can be used in predicting hive population. 

The UrBAN dataset was gathered over the period spanning 2021 to 2022, originating from observations made across a network of ten beehives, part of an urban apiary, positioned in the rooftop of a building located in Montreal, Canada. Various parameters such as audio recordings, temperature measurements, and relative humidity readings were collected. Notably, the dataset encompasses a broad spectrum of critical metrics pertaining to hive inspections, including assessments of colony honey bee population dynamics, evaluations of queen-related conditions, and records of overall health status. Specific health indicators such as \textit{Varroa} mite infestation rates and assessments of winter survivability are also cataloged, providing invaluable insights into the factors influencing hive health and resilience.

\section*{Methods}
\subsection*{Urban Apiary}
\begin{figure}
\centering
\subfloat[]{\label{fig:rooftop}
\includegraphics[angle=-90, width=0.35\linewidth]{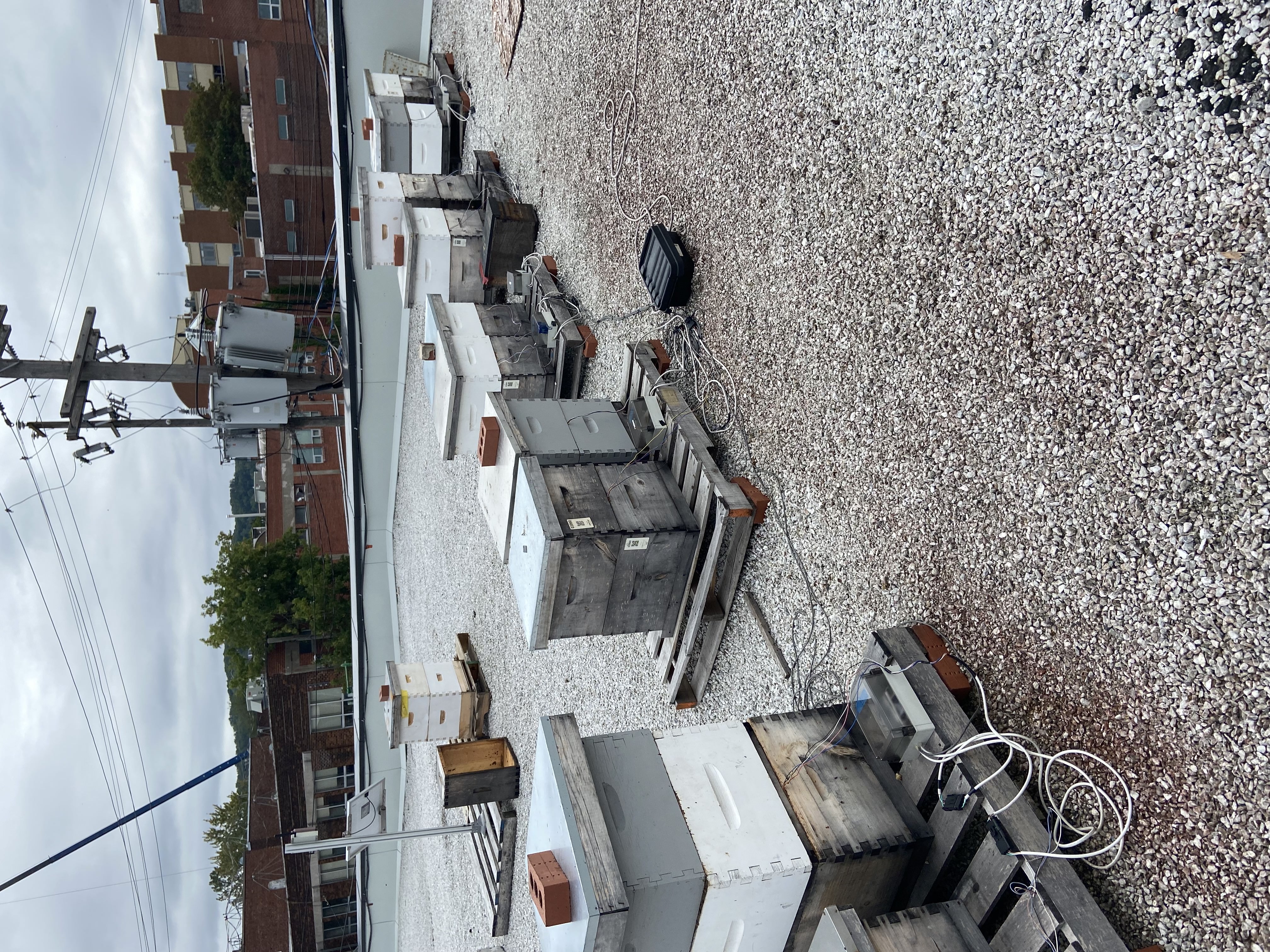}
}
\subfloat[]{\label{fig:audio}
\includegraphics[angle=-90, width=0.35\linewidth]{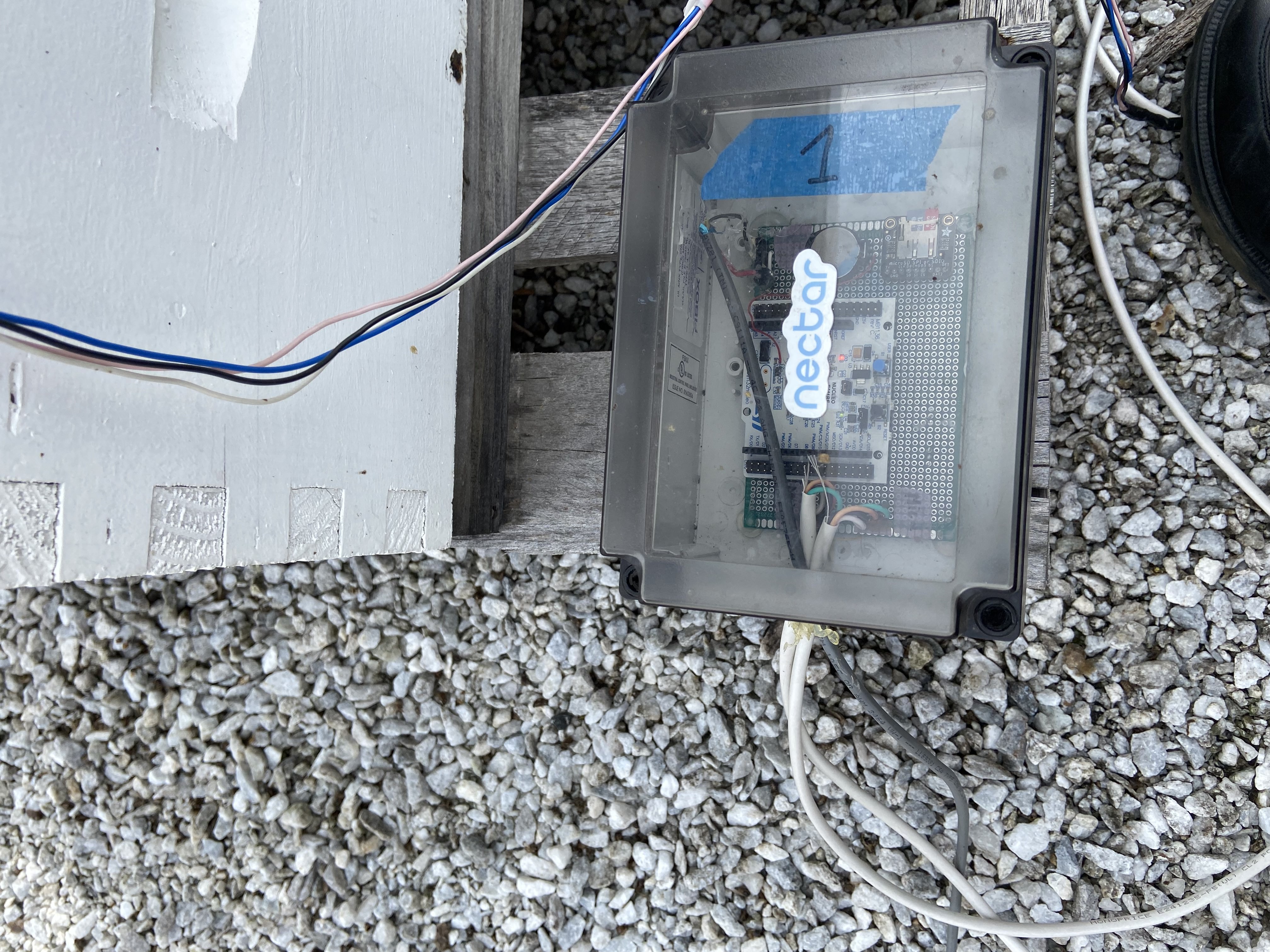}
}
\caption{Photos of the (a) rooftop urban apiary in Montreal and (b) audio recording hardware.}
\label{fig:audio_hardware}
\end{figure}

The rooftop apiary situated in Montréal, Canada (45.5253$^{\circ}$ N, 73.6123$^{\circ}$ W), comprised ten active honeybee hives. These hives were placed on wooden pallets (2 hives per pallet) in a row facing south east. Figure~\ref{fig:rooftop} shows the apiary placement. Each hive consisted of one brood chamber and one to two honey supers, housed within 10-frame standard Langstroth boxes, with a maximum of three boxes per hive. All hives originated from four-frame nucs, which were acquired and installed during the month of May in the year 2021.

Experienced beekeepers deliberately maintained varied hive populations to capture diverse data points, aligning with the primary objective of predicting beehive strength, which correlates with population size. At the beginning of the data collection, each hive contained a distinct number of (full) frames of bees with a minimum of six frames of bees in hives with a single brood box to a maximum of 20 frames in hives with both a brood box and a honey super. As colony populations expanded over time, additional honey supers were introduced. Consequently, within our apiary, the maximum configuration comprised three boxes and up to 30 frames of bees.

\subsection*{Hive Management and Inspection}
\begin{figure}
\centering
\subfloat[]{\label{fig:fob_2021}
\centering
\includegraphics[width=0.49\linewidth]{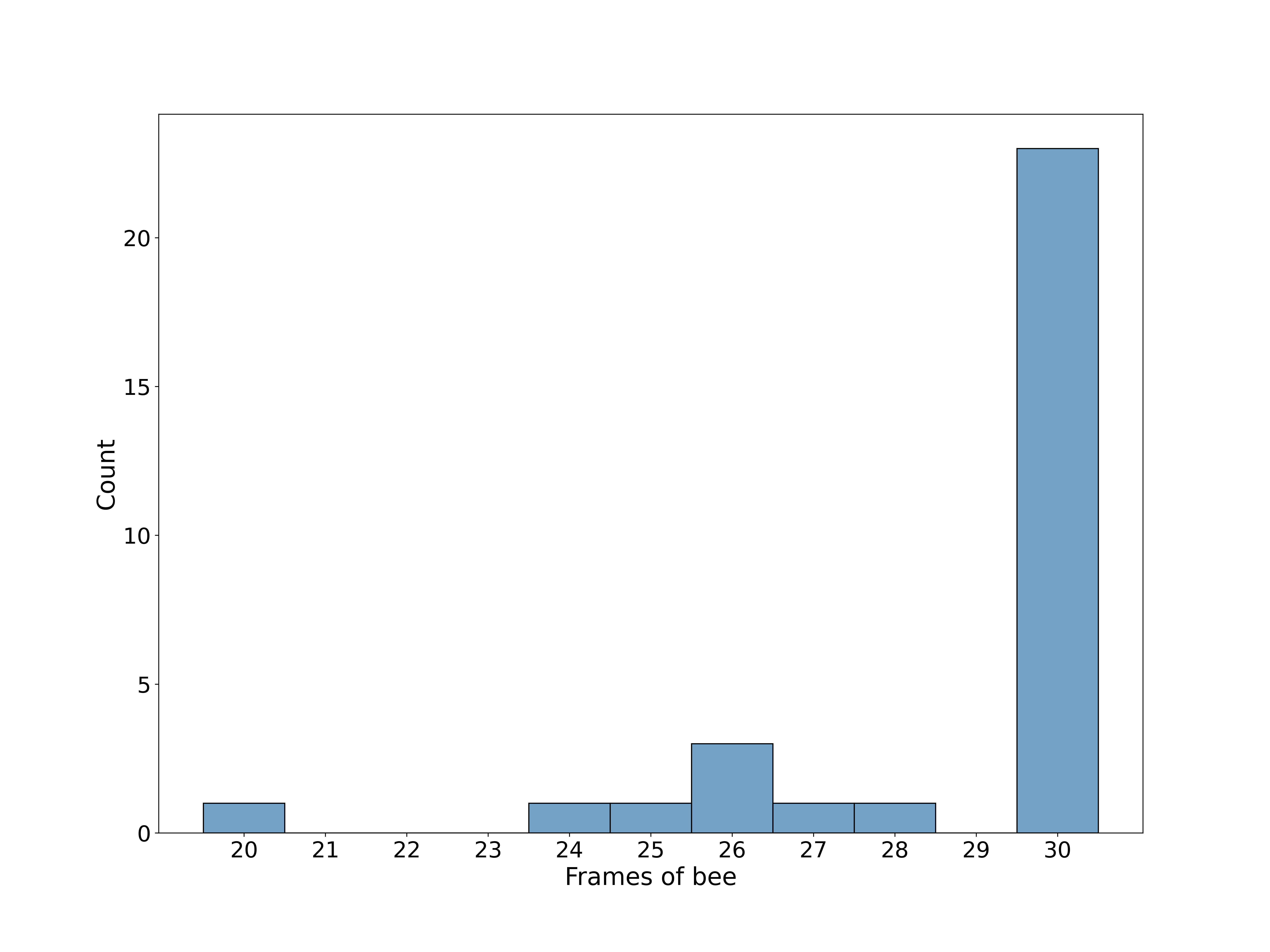}
}
\subfloat[]{\label{fig:fob_2022}
\centering
\includegraphics[width=0.49\linewidth]{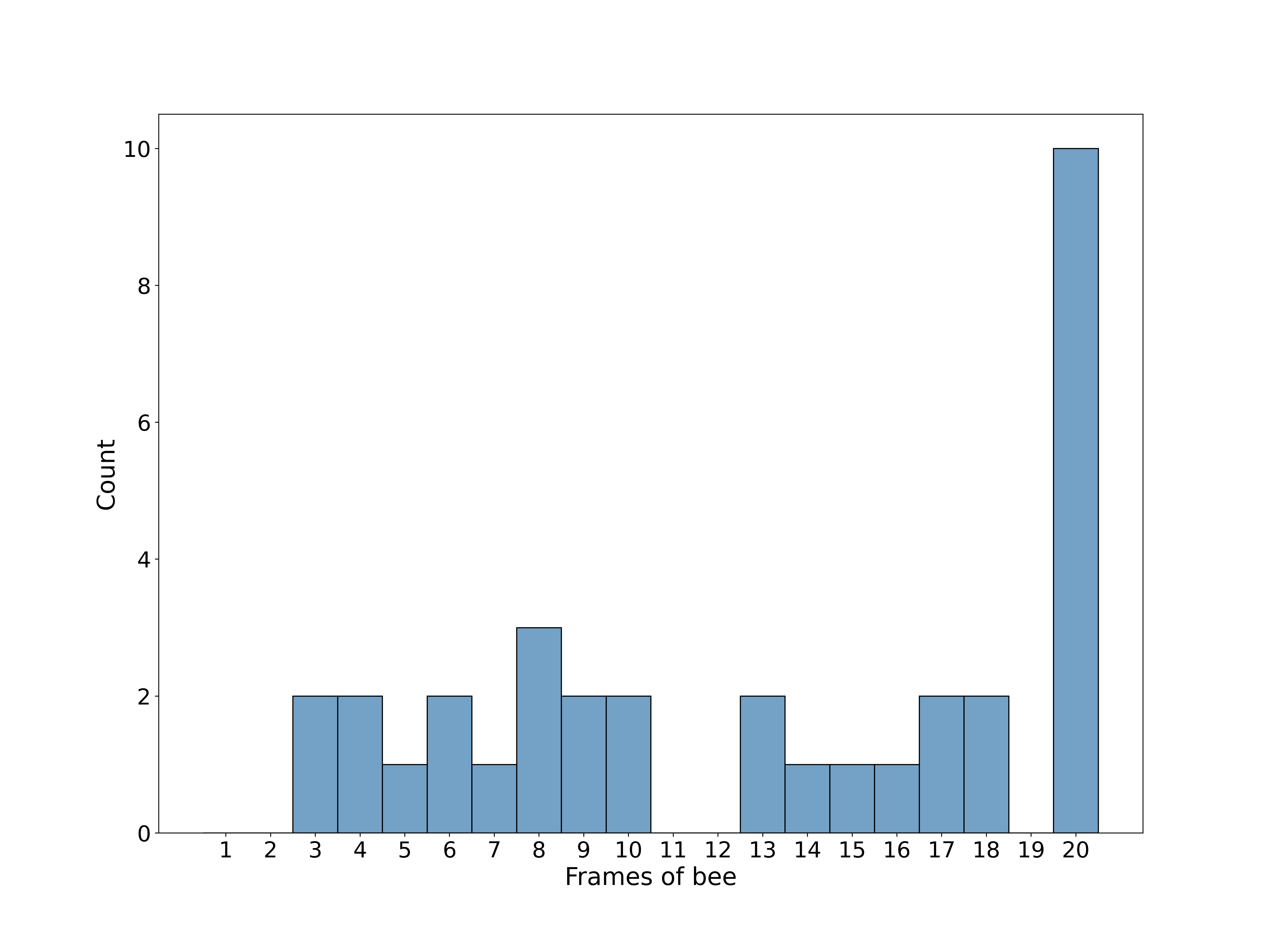}
}
\caption{Histograms of the the number of frames of bees for the year of (a) 2021 and (b) 2022 experiments.}
\label{fig:fob}
\end{figure}

\begin{figure}
\centering
\subfloat[]{\label{fig:fob_2021_detaied}
\centering
\includegraphics[width=0.85\linewidth]{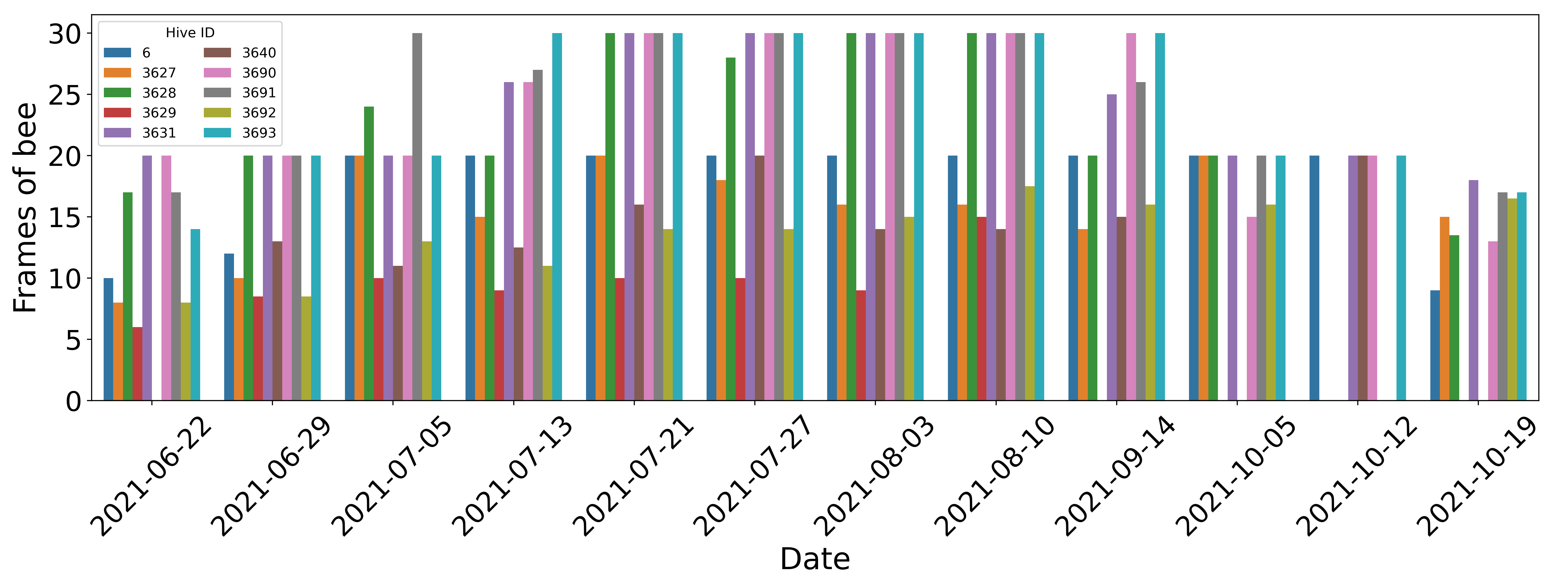}
}

\subfloat[]{\label{fig:fob_2022_detaied}
\centering
\includegraphics[width=0.85\linewidth]{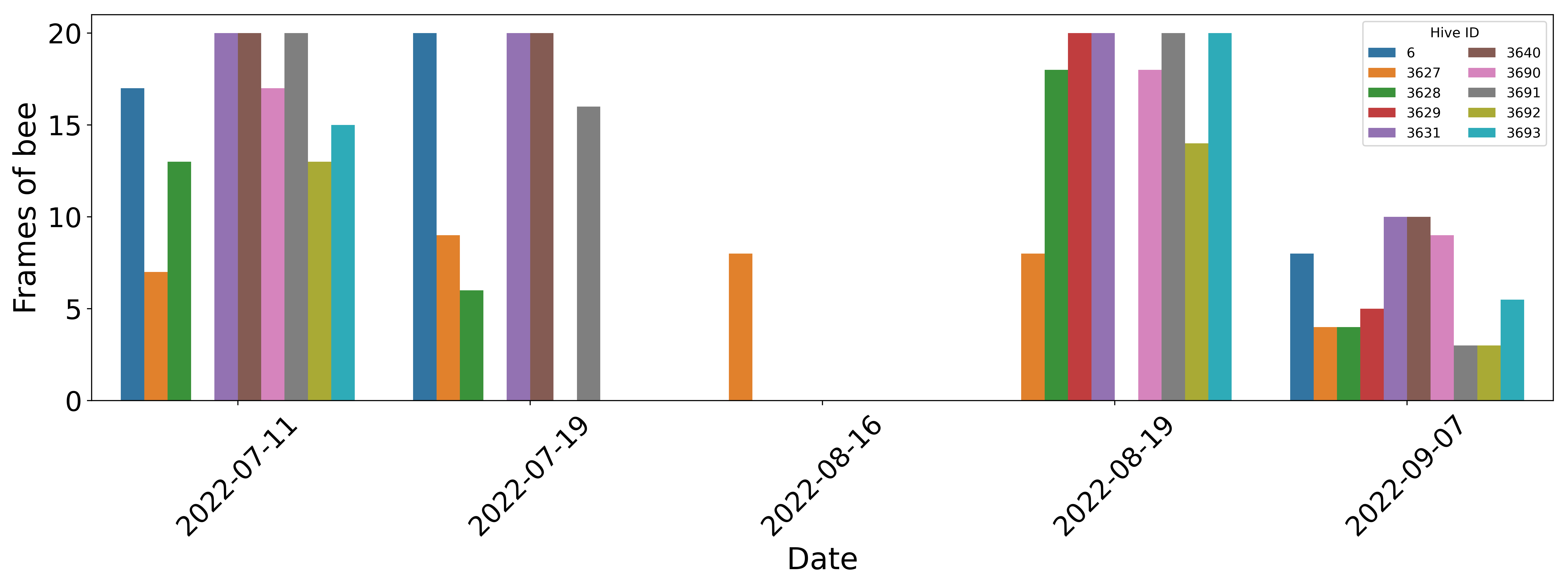}
}
\caption{Barplots of the number of frames of bees on days of inspections for the year of (a) 2021 and (b) 2022 experiments.}
\label{fig:fob_detailed}
\end{figure}

\begin{figure}
\centering
\includegraphics[width=0.4\linewidth]{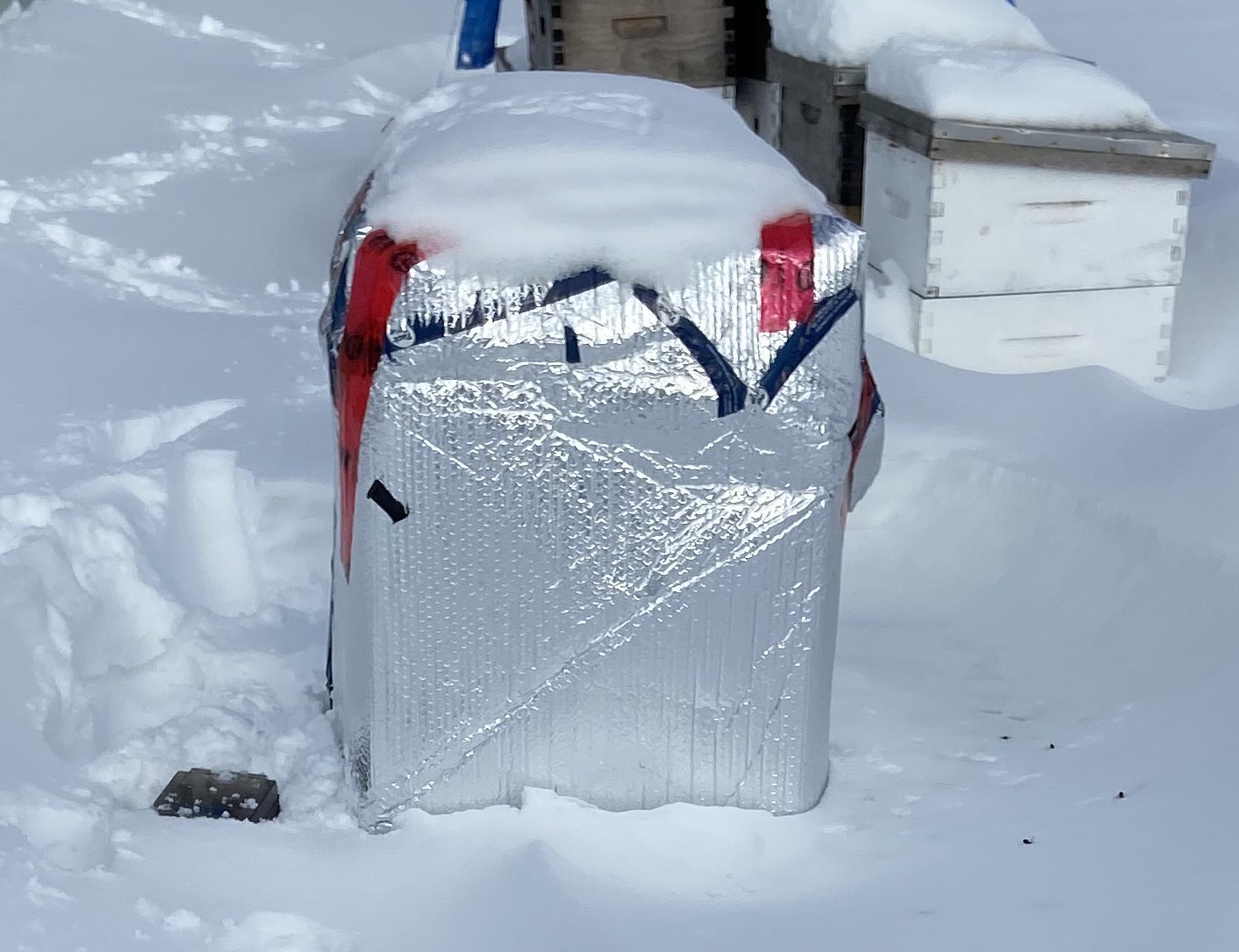}
\caption{Insulation used for beehives overwintering.}
\label{fig:winterization}
\end{figure}

The hives were manually inspected roughly every two weeks to measure the strength of the hives (i.e., the number of frames of bees), to verify the presence of a laying queen, as well as to report any additional observations related to the colony activity. The start and date of these inspections are listed in Table~\ref{tab:hive_management} for each year. Figure~\ref{fig:fob} shows the histogram of the number of frames of bees observed during multiple inspections for each year of the experiments. Each histogram indicates the count of observed frames of bees for all of the beehives during the experiments. Moreover, Figure~\ref{fig:fob_detailed} shows the number of frames of bees for each hive during inspections. 
The beehive colonies were wintered outdoors with the aid of insulation, a strategic approach aimed at enhancing their survivability and well-being during the colder months, as shown in Figure~\ref{fig:winterization}. Insulation provides an additional layer of protection against harsh environmental conditions, helping to maintain stable temperatures within the hive and reduce heat loss.

\begin{table}[]
    \centering
    \begin{NiceTabular}{c|c|c|c|c}
        \toprule
        Year & Inspections  & Varroa mite measurment & Audio recording &  Temperature/Humidity\\
        \midrule
        2021 & \makecell{June 22th\\ - \\October 19th}  & - & \makecell{August 11th\\ - \\October 31st} & \makecell{June 19th\\ - \\October 31st}\\
        \midrule
        2022 & \makecell{July 11th\\ - \\September 7th}  & \makecell{August 24th, \\September 1st,\\ and September 30th} & \makecell{February 1st\\ - \\October 31st} & -\\
        \bottomrule
    \end{NiceTabular}
    \caption{Summary of the data collection start and end dates for each year.}
    \label{tab:hive_management}
\end{table}

\subsection*{Varroa Mite Infestation}
\begin{figure}
\centering
\subfloat[]{\label{fig:varroa_1}
\centering
\includegraphics[width=0.33\linewidth]{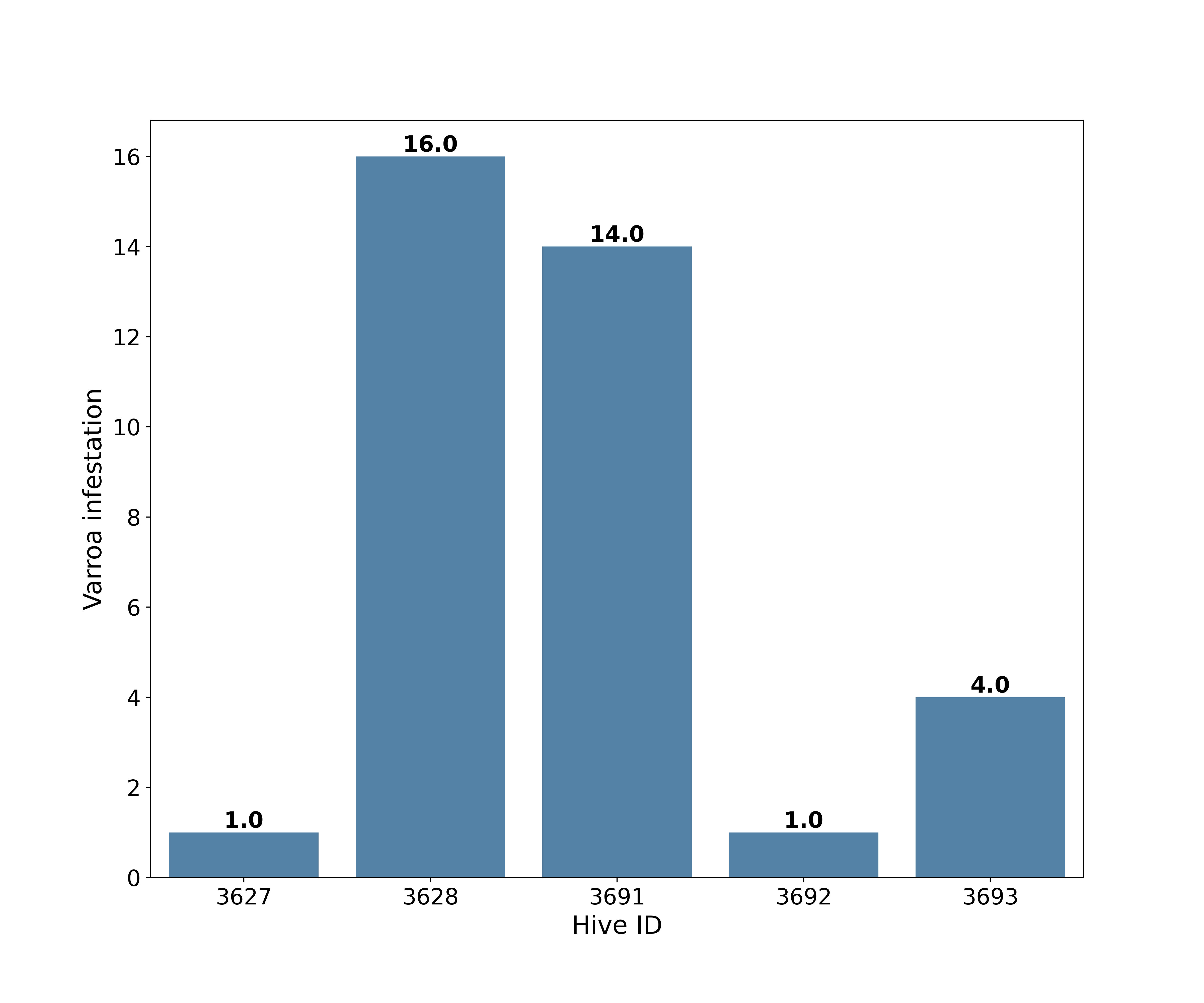}
}
\subfloat[]{\label{fig:varroa_2}
\centering
\includegraphics[width=0.33\linewidth]{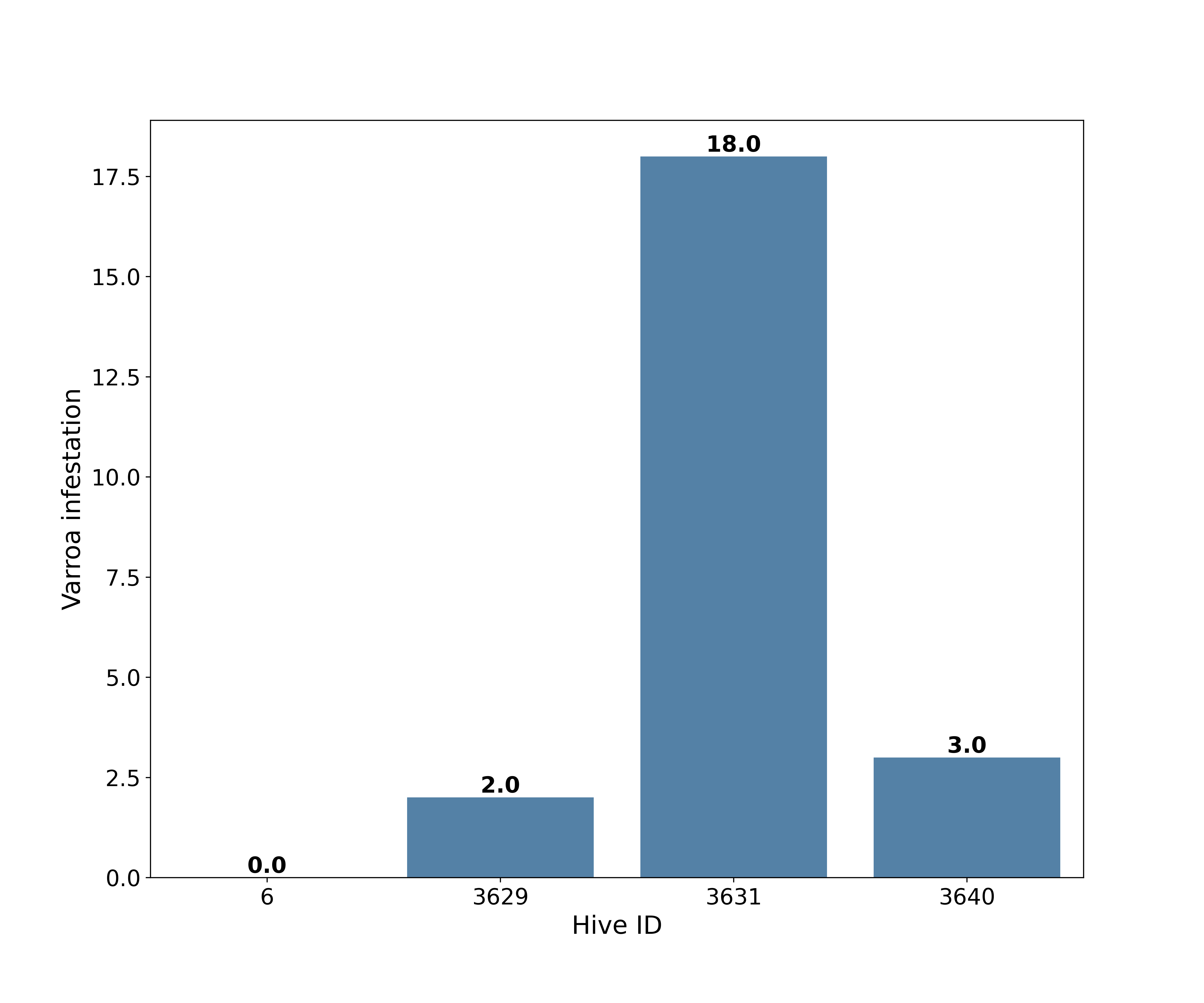}
}
\subfloat[]{\label{fig:varroa_3}
\centering
\includegraphics[width=0.33\linewidth]{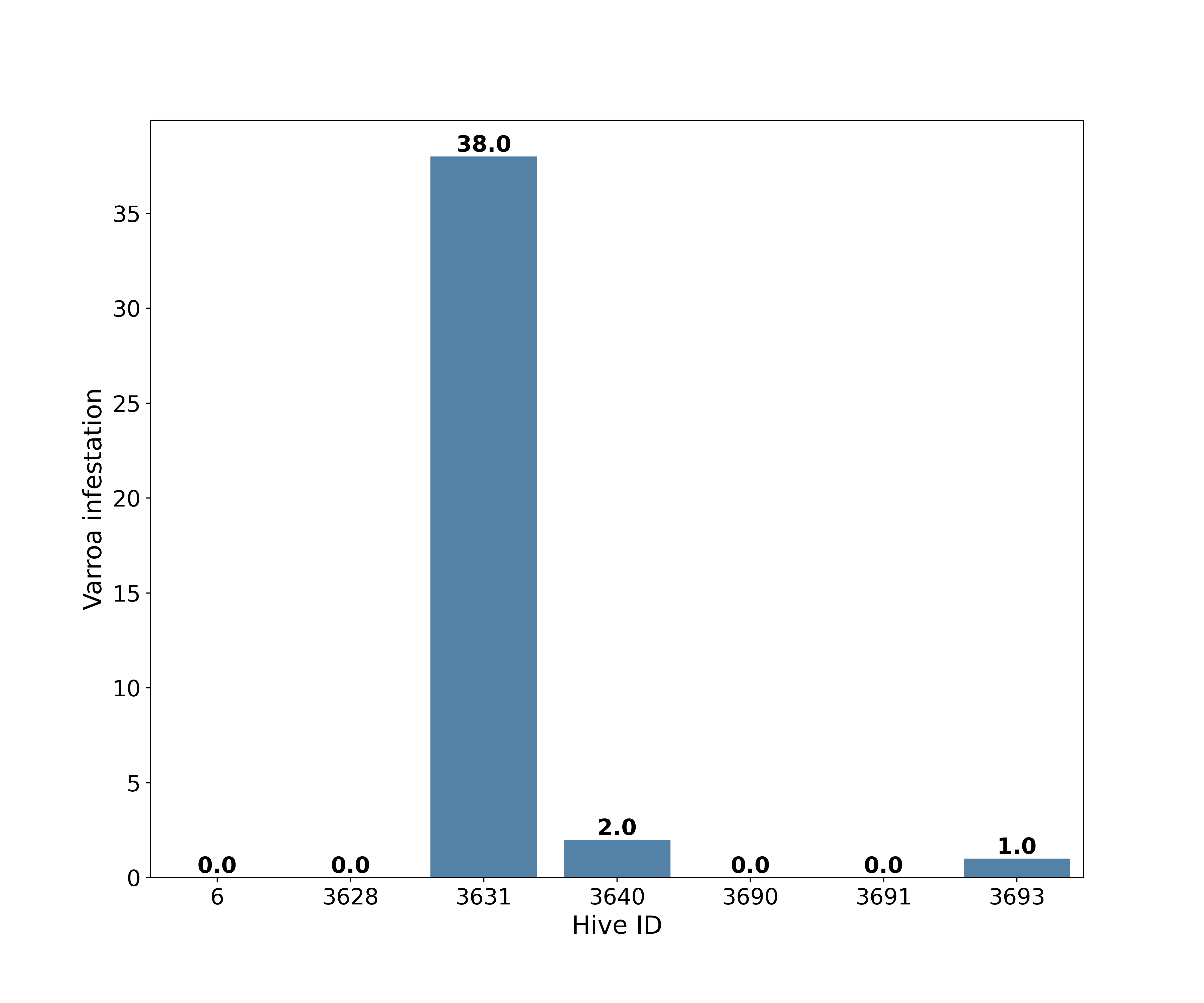}
}
\caption{Varroa mite infestation on (a) August 24th, (b) September 1st, and (c) September 30th, 2022.}
\label{fig:varroa}
\end{figure}

Varroa mite infestation poses a significant threat to the health and well-being of honeybee colonies, making it a matter of utmost importance for beekeepers and researchers alike. In the 2022 data collection, the beekeepers measured the varroa mite infestation using alcohol wash method~\cite{dietemann2013standard} in some of the beehives. The alcohol wash method is a common technique for measuring Varroa mite infestations in honeybee colonies. It involves collecting a sample of about 300 bees from the brood nest, submerging them in isopropyl alcohol (70\% or higher), and gently shaking the container to dislodge the mites. The mixture is then strained, and the mites are counted to estimate the infestation rate. This method, while resulting in the loss of the sampled bees, provides an accurate assessment of mite levels, helping beekeepers make informed decisions about mite control and ensuring colony health. Figure~\ref{fig:varroa} indicates the amount of varroa mite infestation in each measurement. 

\subsection*{Mortality Rate}
Winter mortality is one of the primary causes of beehive losses globally and is a significant concern for both beekeepers and researchers~\cite{kohl2023parasites}. As temperatures drop and resources become scarce, honeybee colonies face numerous challenges that can impact their survival. Factors such as disease prevalence, mite infestations, inadequate food stores, and harsh environmental conditions all contribute to increased mortality rates among bee colonies during the winter~\cite{kempersstatement}. Understanding and monitoring these mortality rates is crucial for assessing the health of bee populations and implementing strategies to mitigate losses. During the data collection, 20\% of the beehives died after overwintering in 2022. 

\subsection*{Sensor Data}
\begin{figure}
\centering
\subfloat[]{\label{fig:sensor}
\centering
\includegraphics[width=0.35\linewidth]{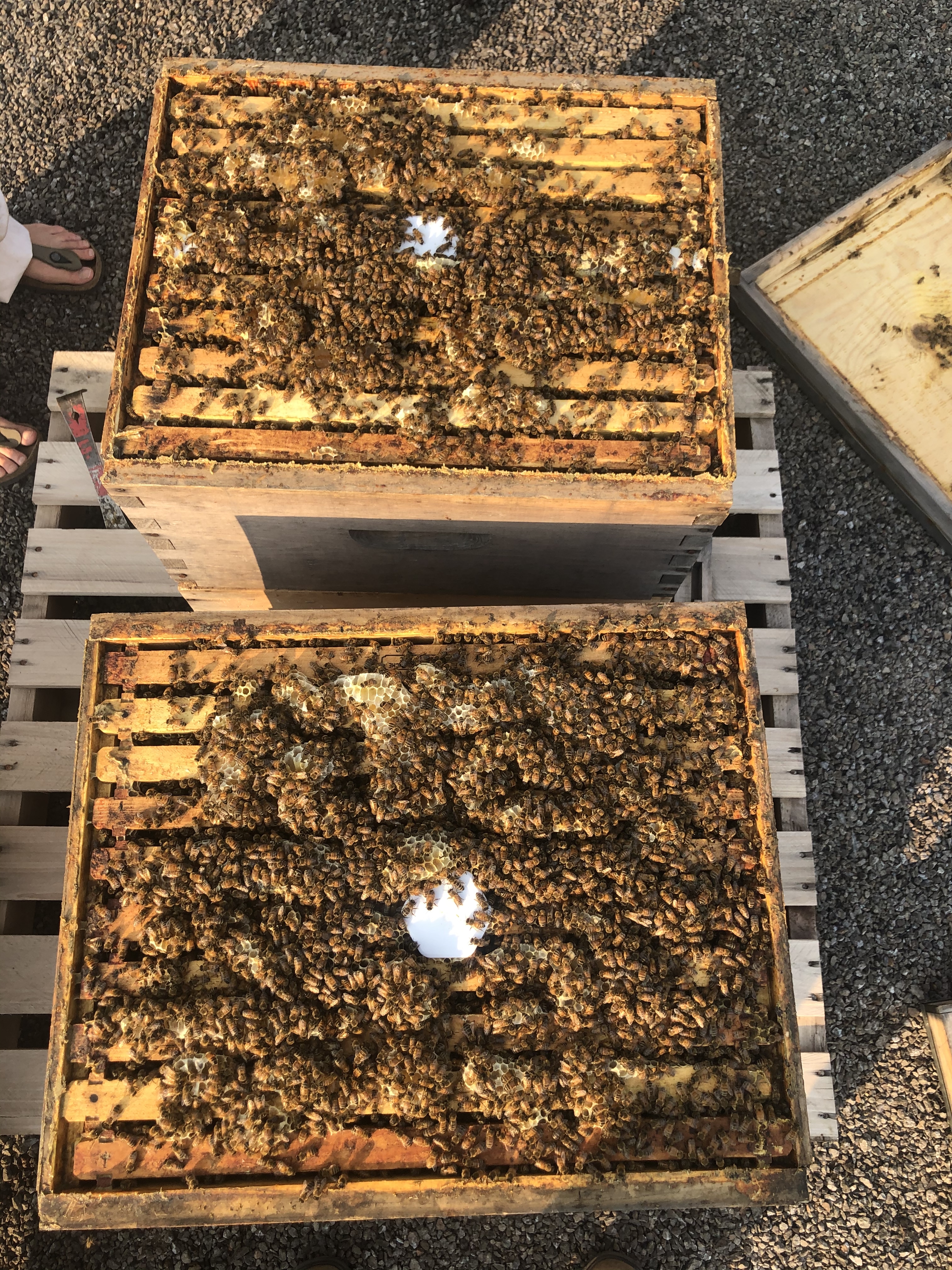}
}
\subfloat[]{\label{fig:microphone}
\centering
\includegraphics[width=0.35\linewidth]{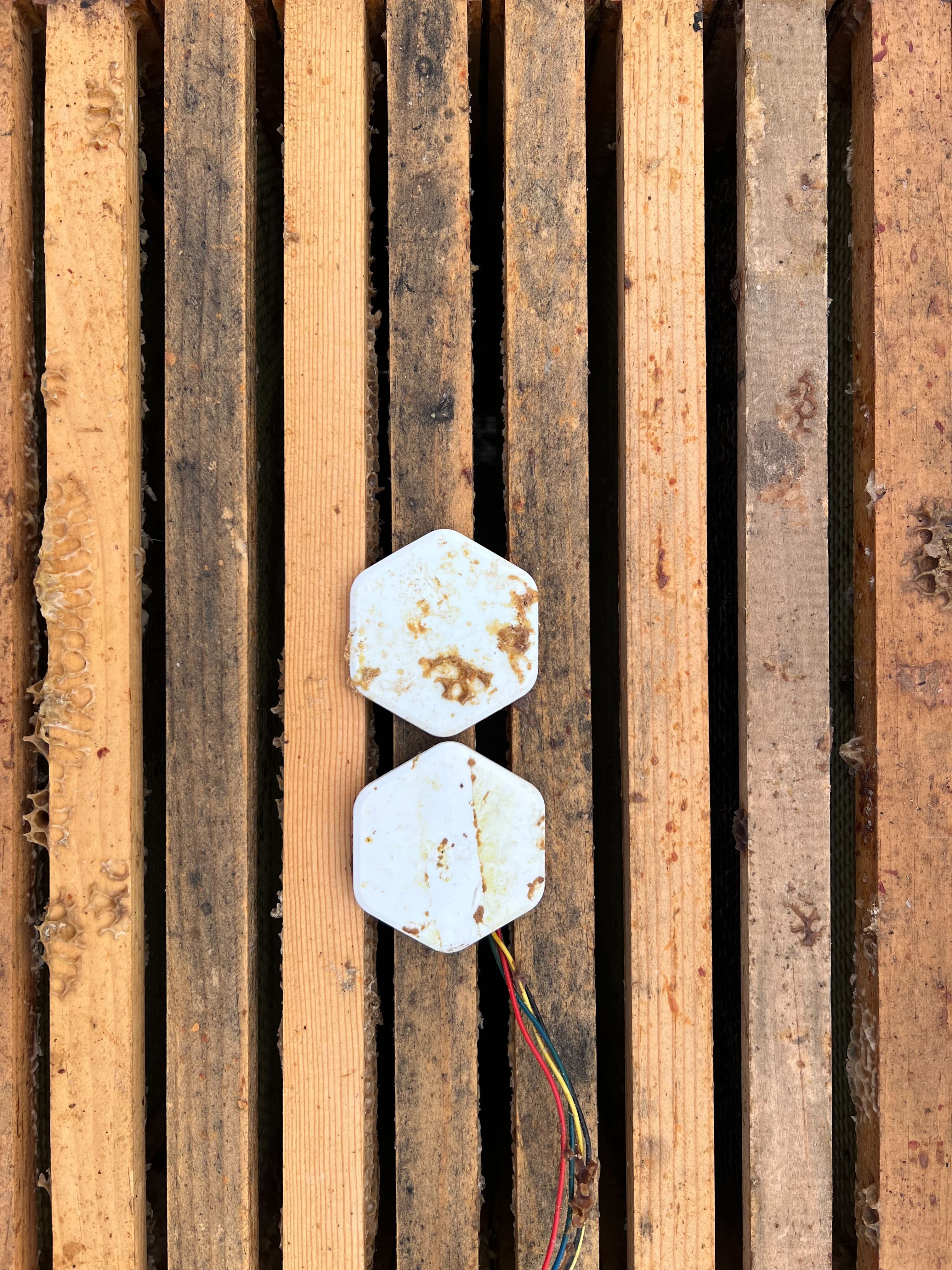}
}
\caption{Location of the (a) temperature and humidity sensor placed on top of the middle frame of the first brood chamber and (b) microphone placed next to it.}
\label{fig:setup}
\end{figure}

A multimodal sensor, positioned atop the central frame within the bottom brood box, facilitated continuous monitoring of internal hive temperature and humidity (Beecon, Nectar Technologies Inc, Canada~\cite{nectar}). Additionally, an accompanying microphone, depicted in Figure~\ref{fig:microphone}, was installed adjacent to the sensor. The multi-modal data is comprised of the average temperature and humidity readings every 15 minutes, and a 15-minute audio segment every 30 minutes with a sampling rate of \SI{48}{\kilo\hertz}. To minimize storage usage, every audio file undergoes resampling to a frequency of \SI{16}{\kilo\hertz}.

\begin{figure}
\centering
\subfloat[]{\label{fig:temp_humid_24h}
\centering
\includegraphics[width=0.5\linewidth]{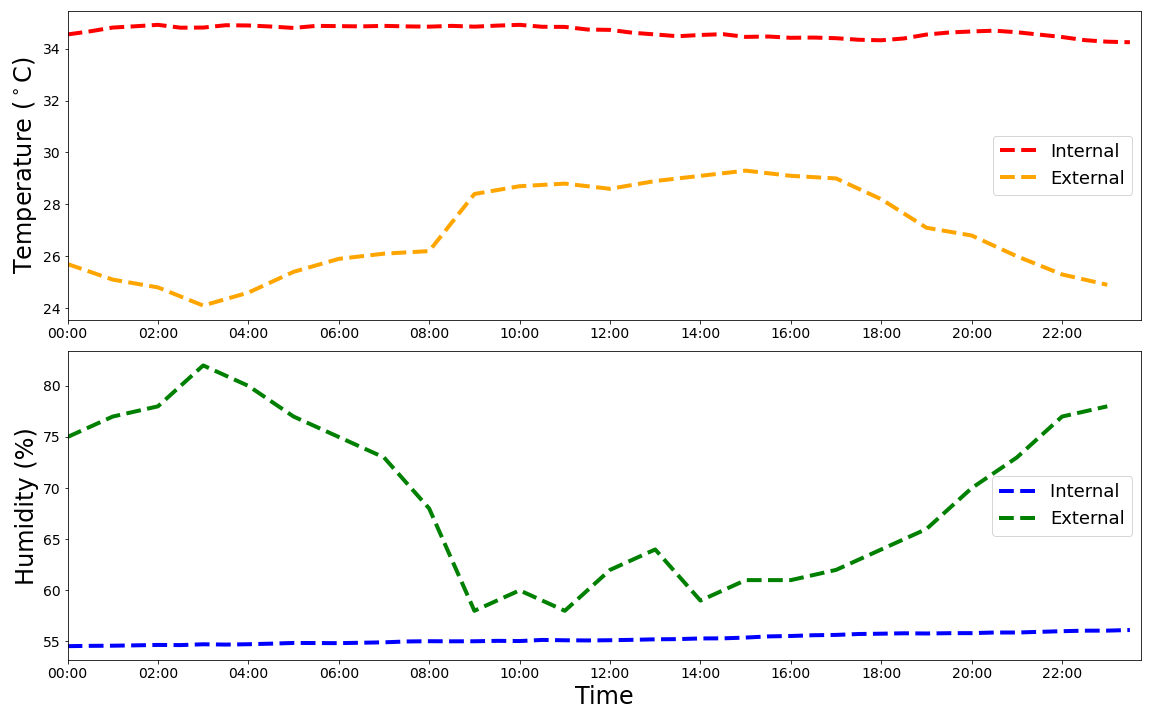}
}
\subfloat[]{\label{fig:temp_humid_total}
\centering
\includegraphics[width=0.5\linewidth]{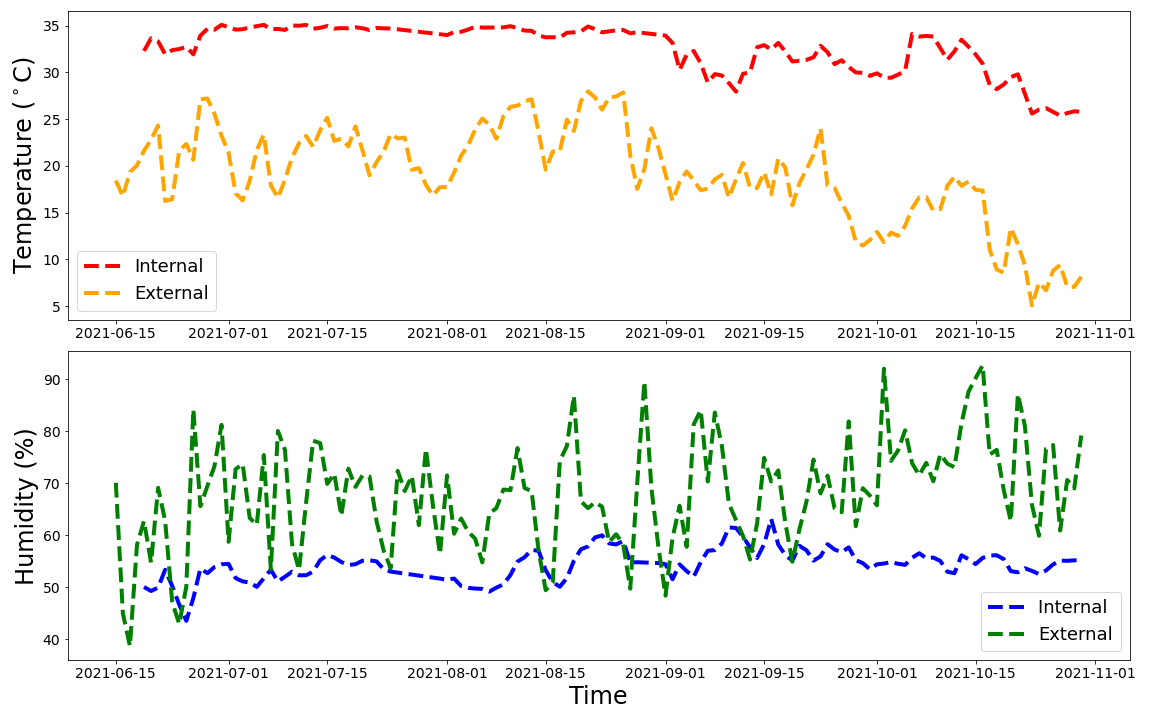}
}
\caption{Internal and external temperature and humidity during (a) a \SI{24}{h} period for a single beehive on August 12th, 2021, and b) total duration of the experiment in 2021.}
\label{fig:temp_humid}
\end{figure}

Moreover, local external temperatures, humidity, and rainfall amounts levels were obtained from the Environment and Climate Change Canada website \footnote{\href{https://climate.weather.gc.ca/}{https://climate.weather.gc.ca/}}. 
A representative example of a \SI{24}{h} snapshot of the changes in internal/external temperature and humidity levels for a single beehive, as well as a 4-month period average of all beehives is shown in Figure~\ref{fig:temp_humid}a and b, respectively. The \SI{24}{h} snapshot (~\ref{fig:temp_humid_24h}) is for a strong and healthy colony in August with one brood chamber and 2 honey supers with a total of 30 frames of bees (covered with at least 70\% of bees). 

\begin{figure}
\centering
\includegraphics[width=0.8\linewidth]{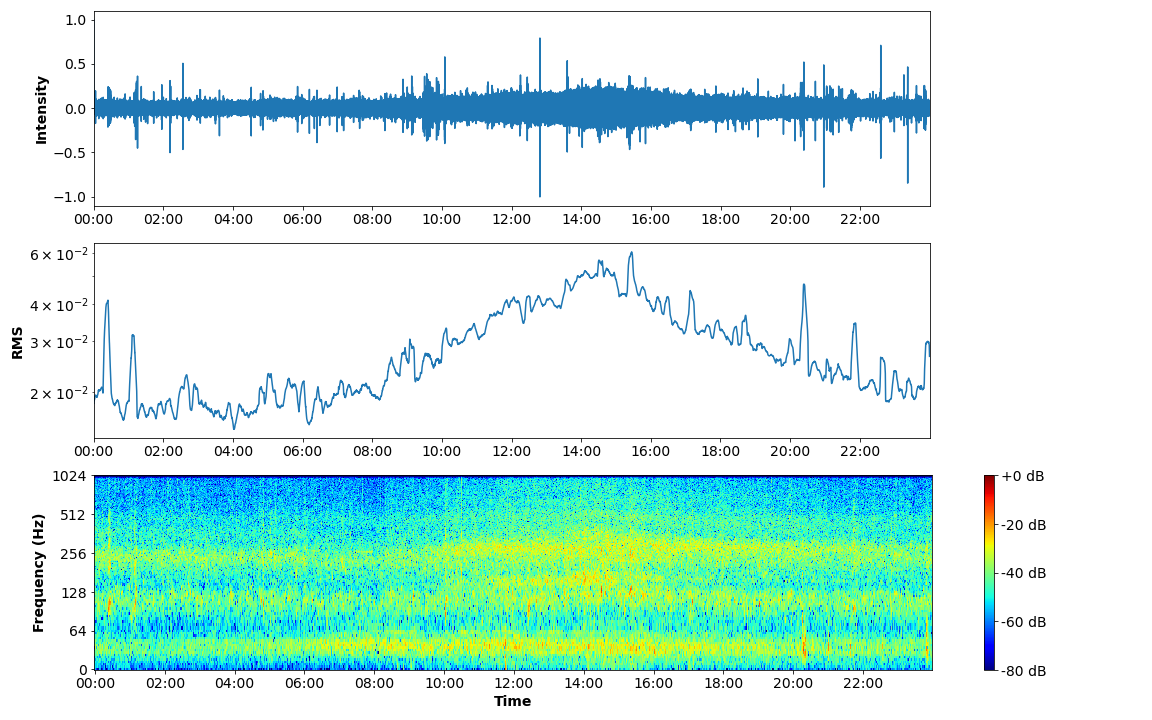}
\caption{Audio intensity, audio RMS, and corresponding spectrogram for a hive with with 3 full boxes of bees (August
12, 2021).
}
\label{fig:audio_rms_spectrogram}
\end{figure}

Figure~\ref{fig:audio_rms_spectrogram} illustrates the intensity of the audio, the smoothed root mean square (RMS) power, and its corresponding spectrogram. It is evident from the figure that the audio power experiences a rise throughout the day, particularly during periods characterized by rising external temperatures and declining humidity levels. This observation suggests increased foraging activity and thermohygrometric regulation within the colony. 

\begin{figure}
\centering
\subfloat[]{\label{fig:audio_barplot_24h_10_20}
\centering
\includegraphics[width=0.33\linewidth]{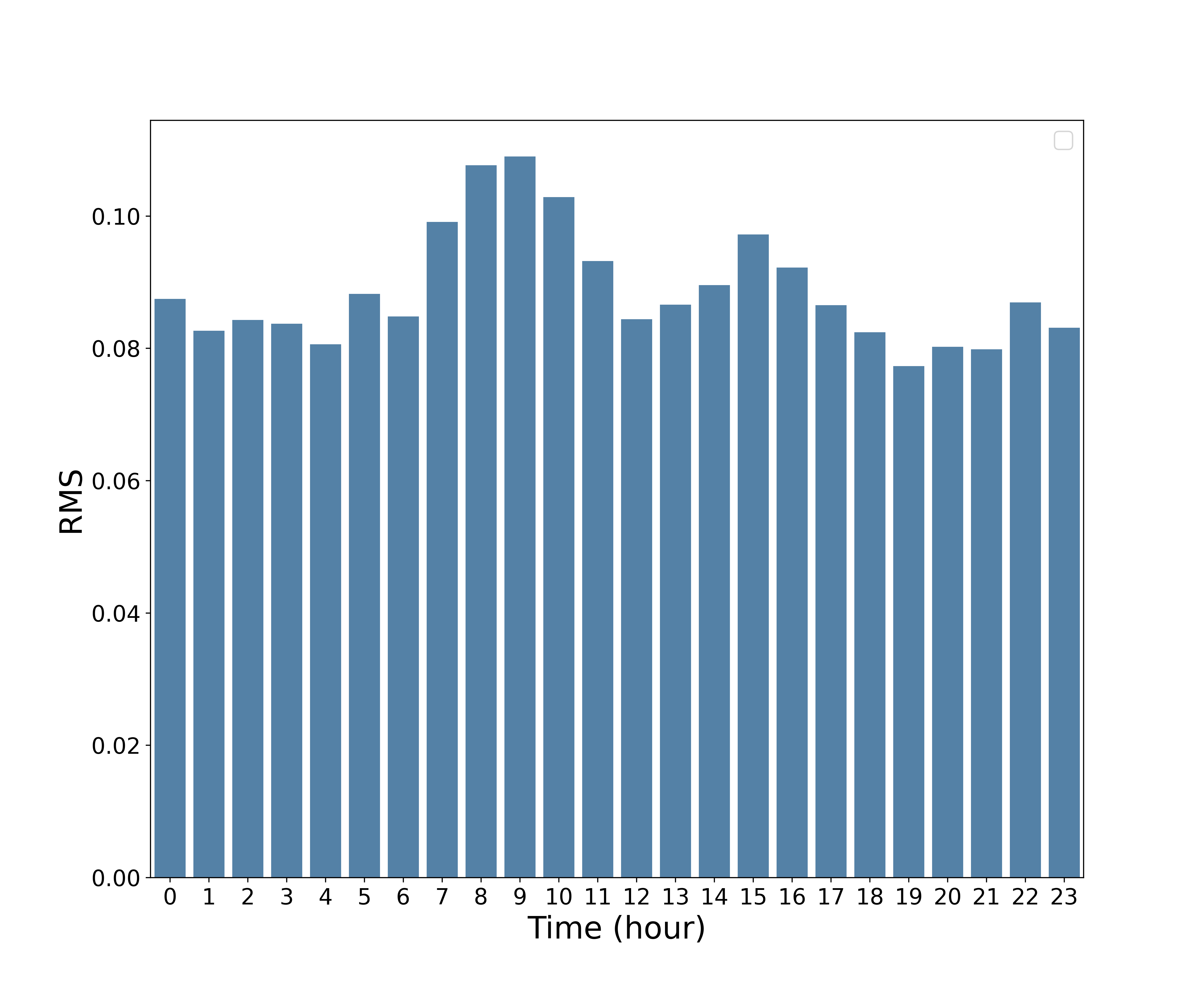}}
\subfloat[]{\label{fig:audio_barplot_24h_20_30}
\centering
\includegraphics[ width=0.33\linewidth]{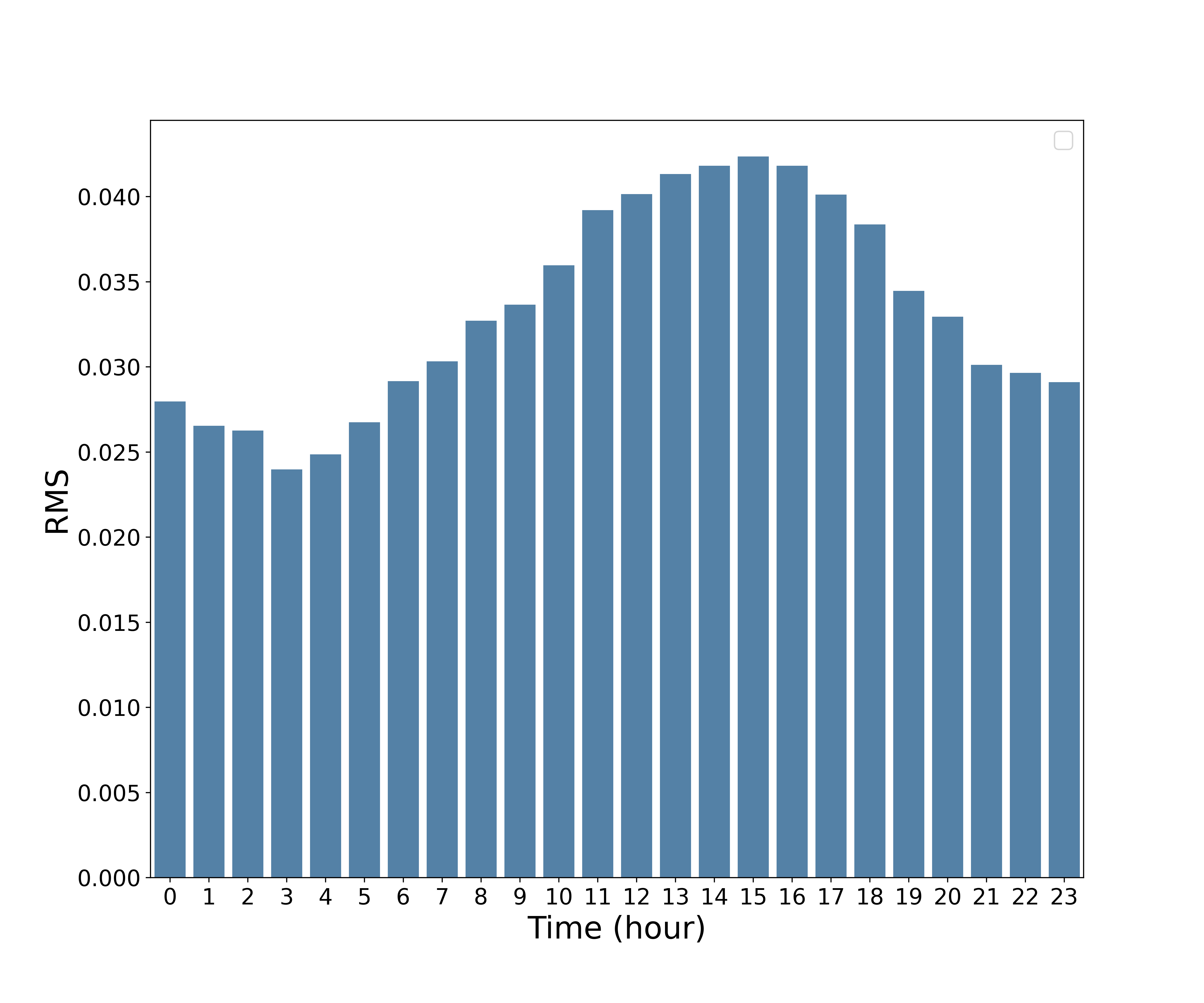}
}
\subfloat[]{\label{fig:audio_barplot_24h_1_30}
\centering
\includegraphics[ width=0.33\linewidth]{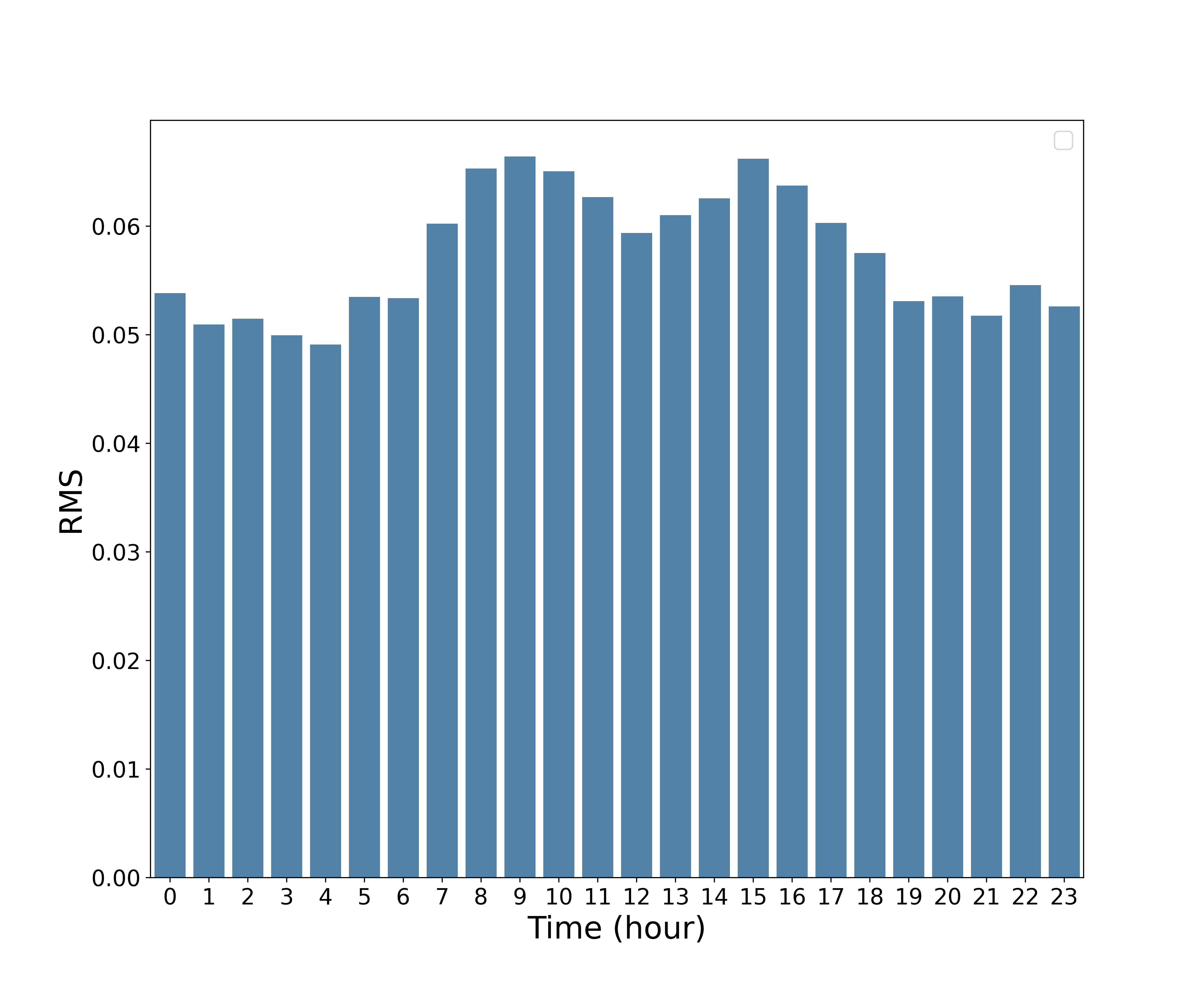}
}
\caption{24h average of RMS value when the number frames of bees is between (a) 10 and 20, (b) 20 and 30, and (c) 1 and 30. }
\label{fig:audio_barplot_24h}
\end{figure}

On the importance of beehive size and its effect on audio, Figure~\ref{fig:audio_barplot_24h} shows the  \SI{24}{h} bar-plots for different number of frames of bees. Each of these plots show the average RMS value with specified frames of bees. Similar to Figure~\ref{fig:audio_rms_spectrogram}, a rising trend during the day can be seen. 
Table~\ref{tab:audio-summary} provides an overview of the quantity and size of raw audio collected for each year. It consists of two columns detailing the total duration in hour and size in gigabytes (\SI{}{\giga\byte}) of recordings, with one column encompassing all recordings regardless of inspection periods, and the other focusing solely on recordings made during the inspection periods specified in Table~\ref{tab:hive_management}.

\begin{table}
\centering
\begin{NiceTabular}{c|c|c|c}
\toprule
Year & Type & Total quantity - size  & Quantity during months of inspections \\
\midrule
2021 & Raw audio & 1752 h  - 185 Gb & 1466 h - 155 Gb\\
\midrule
2022 & Raw audio & 12491 h - 1315 Gb & 1705 h - 180 Gb\\
\bottomrule
\end{NiceTabular}
\caption{The quantity of raw audio for each year of the experiments. }
\label{tab:audio-summary}
\end{table}

\section*{Data Records}
Tables \ref{tab:inspections-summary} and \ref{tab:sensor-summary} provide a comprehensive overview of the inspections (labels) and sensor data, respectively. The inspection files called \url{inspections_2021.csv} and \url{inspections_2022.csv} contain details such as the count of frames of bees, presence of varroa mite infestation, queen status, and mortality rates stored as \url{.csv} files. Each beehive is identified by a unique hive ID, enabling discrimination between them. In Table~\ref{tab:sensor-summary}, the details about recording of internal temperature, humidity, raw audio, and also weather information such as external temperature, humidity, and amount of precipitation are described. The raw audio recordings are \url{wav} files stored in a compressed format for easy download. The sensor data and weather information comprise two \url{.csv} files. 
The UrBAN dataset is made fully publicly available at the Federated Research Data Repository~\cite{urban} (\url{https://doi.org/10.20383/103.0972}). 

\begin{table}
\centering
\begin{NiceTabular}{c|c|c|c}
\toprule
Year & File name &  Columns & Description \\
\midrule
 \multirow{12}{*}{2021} & \multirow{12}{*}{\makecell{\url{inspections_2021.csv}}} & Date & Time stamp (\textsc{YYYY-MM-DD})\\ 
\cmidrule{3-4}
& & Tag number & ID unique to each hive\\
\cmidrule{3-4}
& & Colony size & The number of boxes for each beehive \\
\cmidrule{3-4}
& & Fob 1st & The number of frames of bees in the first box \\
\cmidrule{3-4}
& & Fob 2nd & The number of frames of bees in the second box \\
\cmidrule{3-4}
& & Fob 3rd & The number of frames of bees in the third box \\
\cmidrule{3-4}
& & FoBrood & The number of frames of brood \\
\cmidrule{3-4}
& & Frames of honey & The number of frames of honey \\
\cmidrule{3-4}
& & Queen status & QR/QNS (queen seen or not seen) \\
\cmidrule{3-4}
& & Open & Time stamp indicating opening the box for inspections (\textsc{HH:MM}) \\
\cmidrule{3-4}
& & Close & Time stamp indicating closing the box after inspections (\textsc{HH:MM})\\
\cmidrule{3-4}
& & Note &  Additional observation such as beehive being weak or aggressive\\
\midrule
\multirow{7}{*}{2022} & \multirow{7}{*}{\makecell{\url{inspections_2022.csv}}} & Date & Time stamp (\textsc{YYYY-MM-DD HH:MM:SS})\\
\cmidrule{3-4}
& & Tag number & ID unique to each hive\\
\cmidrule{3-4}
& & Category & \makecell{Hive grading, hive status, frames of bees, varroa, treatment,\\ feeding, custom practice, queen management, hive issues}\\
\cmidrule{3-4}
& & Action detail & \makecell{Detail of each category.\\
Hive grading: 'strong', 'medium', 'weak',\\ 'pulled honey super', 'size - 1d'; \\Hive status: 'queenright', 'queenless', 'deadout';\\ frames of bees: the number of frames of bees; \\Varroa: the varroa mite measurement;\\ Treatment: 'mite away'; \\Feeding: 'sugar';\\ Custom practice: 'add entrance reducer',\\ 'supering', 'added bee escape', 'added trash bag (feeder trick)'; \\
Queen management: 'potential breeder'; \\
Hive issues: 'chalk brood'}\\
\cmidrule{3-4}
& & Queen status & Queenright/queenless \\
\cmidrule{3-4}
& & Is alive & 0/1 (Zero indicates a dead hive)\\
\cmidrule{3-4}
& & Report notes & Additional observation such as beehive being weak or aggressive\\
\bottomrule
\end{NiceTabular}
\caption{Structure of the files describing inspections for each year.}
\label{tab:inspections-summary}
\end{table}

\begin{table}
\centering
\begin{NiceTabular}{c|c|c|c}
\toprule
Year & File/folder name &  Columns & Description \\
\midrule
 \multirow{4}{*}{2021} & \multirow{4}{*}{\makecell{\url{sensor_2021.csv}}} & Date & Time stamp (\textsc{YYYY-MM-DD HH:MM:SS})\\ 
\cmidrule{3-4}
& & Tag number & ID unique to each hive\\
\cmidrule{3-4}
& & Temperature & Internal temperature in degree Celsius\\
\cmidrule{3-4}
& & Humidity & Internal humidity in percentage\\
\cmidrule{2-4}
& \makecell{\url{audio_2021}} & - & \makecell{Audio files names: \\ \texttt{"DD-MM-YYYY\_HHhMM\_HIVE\_Tag.wav"}\\ where 'Tag' is the hive ID number.}\\
\midrule
\multirow{1}{*}{2022} & \multirow{1}{*}{\makecell{\url{audio_2022}}} & - & \makecell{Audio files names:\\ \texttt{"DD-MM-YYYY\_HHhMM\_HIVE\_Tag.wav"}\\ where 'Tag' is the hive ID number.}\\
\midrule
\multirow{5}{*}{2021-2022} & \multirow{5}{*}{\makecell{\url{weather_2021_2022.csv}}} &  Date/Time (LST) & Time stamp (\textsc{YYYY-MM-DD HH})\\
\cmidrule{3-4}
& & Temp (°C) & External temperature in degree Celsius\\
\cmidrule{3-4}
& & Rel Hum (\%) & External humidity in percentage\\
\cmidrule{3-4}
& & Wind Spd (km/h) & The speed of wind\\
\cmidrule{3-4}
& & Precip. Amount (mm) & The amount of precipitation\\
\bottomrule
\end{NiceTabular}
\caption{Structure of the files/folders of the sensor data (temperature and humidity), raw audio recordings, and weather information.}
\label{tab:sensor-summary}
\end{table}

\section*{Technical Validation}
\subsection*{Audio Enhancement}

Removing environmental noise from bee acoustic audio is crucial for enhancing the effectiveness of monitoring systems in beekeeping. Environmental noise can obscure the sounds produced by bees, making it difficult to accurately detect and analyze important behaviors and events within the hive. Moreover, it could reduce the accuracy of the monitoring system significantly. While some studies explored methods in removing noises, such as human speech~\cite{10289037, 10424387}, there is still a need for a general noise removal step. 

\begin{figure}
\centering
\includegraphics[width=0.75\linewidth]{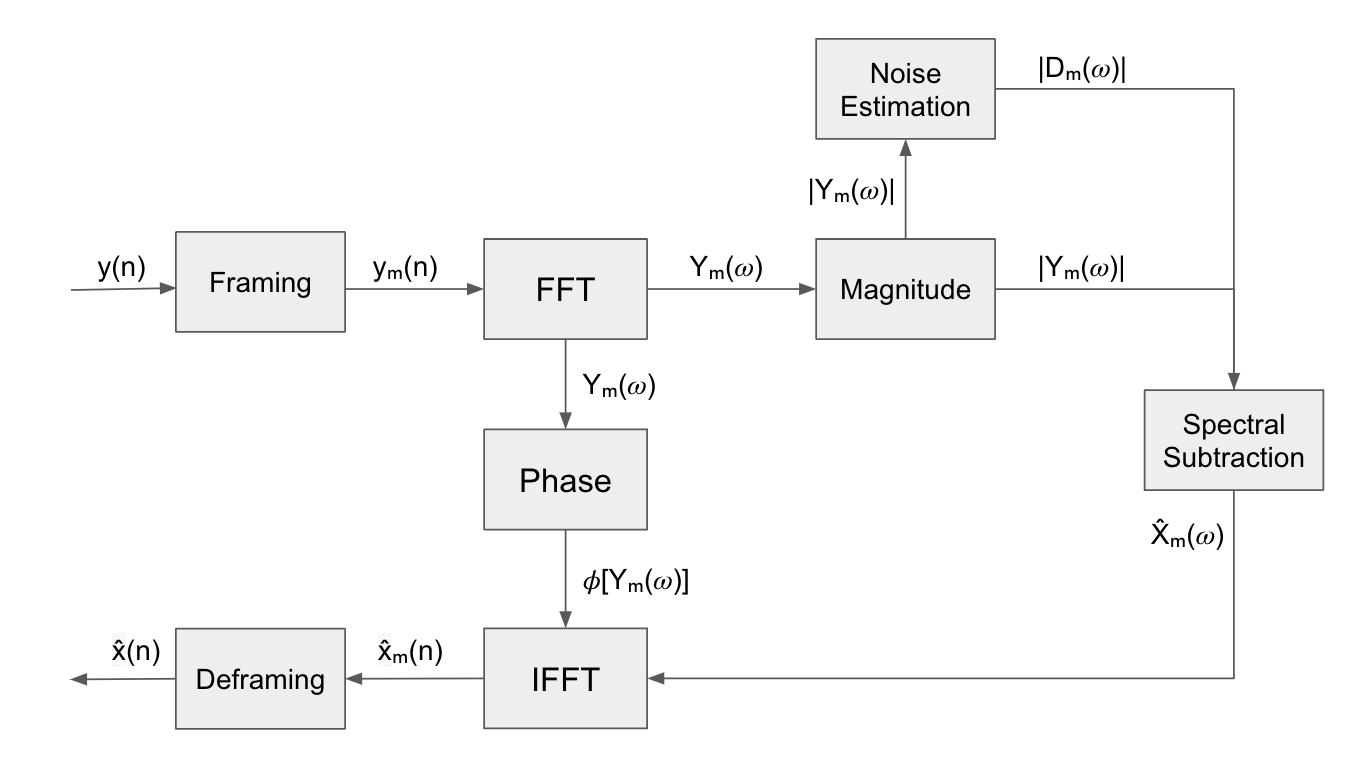}
\caption{Block diagram of the spectral amplitude subtraction.}
\label{fig:spectral_subtraction_diagram}
\end{figure}

Spectral amplitude subtraction is a technique used in audio processing to enhance the quality of audio recordings. As shown in the block diagram in Figure~\ref{fig:spectral_subtraction_diagram}, it involves subtracting the spectral components of noise or unwanted signals from the original audio signal to reduce background noise and improve signal clarity. By identifying the spectral profile of noise or interference and subtracting it from the audio signal, spectral amplitude subtraction helps isolate the desired sound, resulting in cleaner and more intelligible audio recordings. Figure~\ref{fig:spectral_subtraction_diagram} shows the diagram of the spectral subtraction algorithm, where $y(n)$ and $\hat{x}(n)$ are the noisy and cleaned signals, respectively. As the figure indicates, after framing the audio and calculating the fast Fourier transform (FFT), the noise spectral profile needs to be estimated and eventually subtracted from the noisy signal. 

\begin{figure}
\centering
\includegraphics[width=0.85\linewidth]{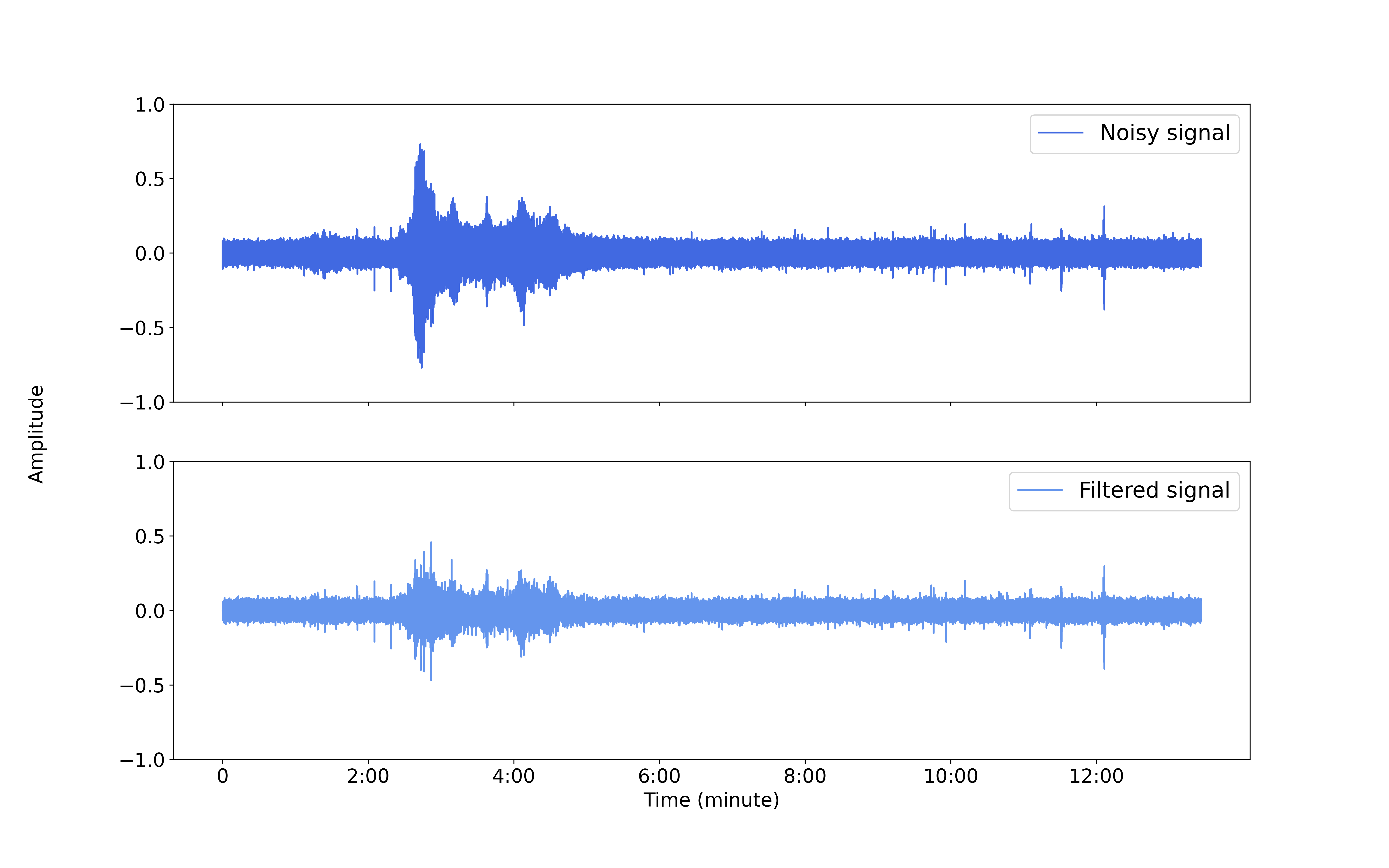}
\caption{From top to bottom, noisy audio amplitude of the beehive, and the corresponding filtered audio.}
\label{fig:spectral_subtraction_example}
\end{figure}

Here, we used an exponential moving average (EMA) filter to estimate the noise. 
\begin{equation}
    d[n] =  (1-\alpha) y[n] +  \alpha d[n-1],
\end{equation}
where $d[n]$ and $\alpha$ are the noise signal and the weighting factor or smoothing parameter, respectively. In order to only update $\alpha$ in noisy frames, an adaptive algorithm is used~\cite{lin2003adaptive}. Using this method, $\alpha$ will be a function of a-posteriori signal-to-noise ratio (SNR), i.e., 
\begin{equation}
    \alpha(m) = \frac{1}{1+exp^-a(SNR(m)-T)},
\end{equation}
where the parameter $a$ represents the steepness of the sigmoid function's slope and the SNR is estimated by $|Y_m(\omega|^2/E[D_(m-N)(\omega)]$, using the $N$ previous frames of audio. The scripts for the spectral subtraction algorithm are made available at at our Github repository (\url{https://github.com/MuSAELab/UrBAN}) to facilitate study replication. Figure~\ref{fig:spectral_subtraction_example} depicts the audio signal collected from the empty hive (top plot), along with one of the populated hives before (middle) and after (bottom) enhancement.

\subsection*{Feature Extraction}
A machine learning framework for predicting hive strength through audio analysis comprises several essential stages, including signal measurement, pre-processing, feature extraction, and regression. Following the enhancement of the audio signal and the removal of unwanted noise, feature extraction becomes crucial. In this process, four distinct feature sets are derived for predicting the state of bee audio frames: mel-frequency cepstral coefficients (MFCCs), linear-frequency cepstral coefficients (LFCCs), spectral shape descriptors, and some hand-crafted parameters described in \cite{zhu2023mspb}.

MFCCs have emerged as a cornerstone in audio-based applications, replicating the auditory processing mechanism of the human ear by employing mel-scale frequency mapping before cepstrum analysis~\cite{davis1980comparison}. Within the realm of precision beekeeping, these features have garnered considerable attention, prominently featured in approximately 30\% of studies examined in previous reviews, particularly for tasks such as bee and queen presence, as well as swarming detection~\cite{abdollahi2022automated}. Here, 12 coefficients are extracted alongside the zeroth coefficient, utilized as a measure of signal power using 26 mel filters. Additionally, LFCCs are extracted through linear filters for comparison with MFCCs.

Spectral shape descriptors play a crucial role in the analysis and characterization of audio signals. In this paper, nine spectral shape descriptors are computed, including centroid, spread, skewness, kurtosis, entropy, rolloff, flatness, crest, and flux.
Furthermore, other works have relied on hand-crafted audio features, including hive power (power between 122 Hz and 515 Hz), audio band density ratio (the ratio of hive power to the power of the entire frequency range), audio density variation (reflecting changes in hive power within each audio frame), and audio band coefficients in 16 linearly spaced frequency bins. For further details, interested readers are directed to~\cite{zhu2023mspb}. After feature extraction, daily statistics (mean, std, skewness, and kurtosis) of all samples were calculated to be used as inputs for different classificiation/regression tasks. Scripts to extract all the features tested herein are made available on our Github repository (\url{https://github.com/MuSAELab/UrBAN}) to facilitate experiment replication.

\subsection*{Validation: Frame of Bees Prediction}
Here, we use frames of bees prediction as a task to validate the dataset under an ML framework. In accordance with recommendations from previous studies~\cite{abdollahi2022importance, ieee_26}, two distinct experimental configurations are explored: "random-split" and "hive-independent". In the random-split approach, the complete dataset spanning 18 (one of the beehives had a problem in audio recording) hives from the years 2021 and 2022 is randomly partitioned into three segments: 25\% for testing, 25\% for validation, and 50\% for training. Conversely, in the hive-independent setup, 10 hives are used for training, 4 for validation, and 4 for testing. This process was repeated for 10 iterations for each split scenario to provide a more robust performance analysis.  

In the domain of ML, feature selection is important to avoid the curse of dimensionality and to improve the generalization abilities of models. By carefully selecting features, redundant or irrelevant attributes can be excluded, thereby reducing the risk of overfitting and improving the model's capacity to identify meaningful patterns within the data. To achieve this goal, several feature selection techniques were explored, including random forest feature importance~\cite{breiman2001random, genuer2010variable}, Principal Component Analysis (PCA)~\cite{mackiewicz1993principal}, minimum Redundancy Maximum Relevance (mRMR)~\cite{peng2005feature}, and SHAP (SHapley Additive exPlanations~\cite{lundberg2017unified}. Each feature set underwent testing with these methods, and the most effective approach was determined based on performance metrics evaluated on the validation set. Subsequently, a random forest regressor was employed to predict the number of frames of bees. Model evaluation was conducted using three key metrics: mean absolute error (MAE), root-mean-square error (RMSE), and Pearson correlation of the predictions relative to the ground truth values.

Table~\ref{tab:perforamance} presents the performance of each feature set both before and after pre-processing/enhancement under the two different split scenarios. To ensure the significance of the results, a random baseline regressor is also employed. Predictions are generated by drawing random numbers within a specified range for each instance in the test set. Each metric is presented as an average $\pm$ standard deviation with its corresponding p-value from the t-test, indicating the statistical significance of the model's performance compared to a random baseline. In the "random-split" case, the MFCCs outperform the baseline model and also other features both with or without pre-processing. In the "hive-independent" case, all features have lower performance compared to the "random-split", while the spectral descriptors achieve the best results. On the importance of audio enhancement and removing noise, the results indicated that in most of the experiments, the performance improved after spectral amplitude subtraction. 

\begin{table}
\centering
\makebox[\textwidth][c]{
            \resizebox{\linewidth}{!}{%
\begin{NiceTabular}{c|cccccc}
\toprule
& \multicolumn{6}{c}{\textbf{Random-Split}} \\
\cmidrule{2-7}
\multirow{2}{*}{Features} &
  \multicolumn{3}{c}{No pre-processing} &
  \multicolumn{3}{c}{Spectral amplitude subtraction} \\
  \cmidrule{2-7}
  & {MAE} & {RMSE} & {Correlation} &  {MAE} & {RMSE} & {Correlation} \\
  \cmidrule{1-7}
 Random baseline & \makecell{4.47 $\pm$  0.35} & \makecell{5.35 $\pm$  0.30} & \makecell{0.70 $\pm$  0.04} & & & \\
LFCCs & \makecell{2.95 $\pm$  0.41***} & \makecell{3.94 $\pm$  0.45***} & \makecell{0.83 $\pm$ 0.05***} & \makecell{2.75 $\pm$  0.95***} & \makecell{3.82 $\pm$  0.91***} & \makecell{0.83 $\pm$  0.06**} \\
MFCCs & \makecell{2.58 $\pm$ 0.88***} & \makecell{3.57 $\pm$ 1.00***} & \makecell{0.87 $\pm$ 0.06***} & \makecell{2.05 $\pm$  0.34***} & \makecell{2.99 $\pm$  0.46***} & \makecell{0.90 $\pm$  0.028***} \\
Spectral descriptors & \makecell{3.30  $\pm$ 0.44***} & \makecell{4.19  $\pm$0.42***} & \makecell{0.80 $\pm$ 0.05**} & \makecell{3.22 $\pm$ 0.62**} & \makecell{4.08 $\pm$  0.63***} & \makecell{0.81 $\pm$  0.06ns} \\
Hand-crafted & \makecell{3.73  $\pm$ 0.46**} & \makecell{4.73 $\pm$0.53**} & \makecell{0.73 $\pm$ 0.06ns} & \makecell{3.48 $\pm$ 0.43***} & \makecell{4.42 $\pm$ 0.43***} & \makecell{0.79 $\pm$ 0.07***} \\
\toprule
& \multicolumn{6}{c}{\textbf{Hive-Independent}} \\
\cmidrule{2-7}
Random baseline & \makecell{4.48 $\pm$  0.59} & \makecell{5.38 $\pm$  0.59} & \makecell{0.72 $\pm$  0.13} & & & \\
LFCCs & \makecell{4.36 $\pm$ 0.80ns} & \makecell{5.33 $\pm$ 0.76ns} & \makecell{0.67 $\pm$  0.10ns} & \makecell{4.01 $\pm$ 1.03**} & \makecell{4.88 $\pm$ 1.00*} & \makecell{0.70 $\pm$ 0.15ns} \\
MFCCs & \makecell{4.18 $\pm$ 0.10ns} & \makecell{4.98 $\pm$ 0.65ns} & \makecell{0.71 $\pm$ 0.10ns} & \makecell{3.98 $\pm$  0.61**} & \makecell{4.93 $\pm$  0.69*} & \makecell{0.70 $\pm$  0.15ns} \\
Spectral descriptors & \makecell{4.07 $\pm$ 0.74*} & \makecell{5.01 $\pm$ 0.67*} & \makecell{0.69 $\pm$ 0.12ns} & \makecell{3.97 $\pm$ 0.48**} & \makecell{4.77 $\pm$ 0.5***} & \makecell{0.78 $\pm$ 0.11**} \\
Hand-crafted & \makecell{4.47  $\pm$0.82ns} & \makecell{5.30 $\pm$ 0.83ns} & \makecell{0.59 $\pm$ 0.20ns} & \makecell{4.27 $\pm$ 0.71†} & \makecell{5.13 $\pm$ 0.69*} & \makecell{0.67 $\pm$ 0.21ns} \\
\bottomrule
\end{NiceTabular}
}
}
\caption{Performance comparison between different feature sets and data partitioning setups, with and without spectral enhancement. The significance of p-values from the t-test is indicated as follows: 
$p<0.001$ is indicated as *** (very significant), 
$p<0.01$ is indicated as ** (significant), 
$p<0.05$ is indicated as * (moderately significant), 
$p<0.1$ is indicated as † (marginally significant), and 
$p \geq 0.1$ is indicated as ns (not significant)}
\label{tab:perforamance}
\end{table}

\section*{Usage notes}
The four CSV files including, \url{inspections_2021}, \url{inspections_2022}, \url{sensor_2021}, and \url{weather_2021_2022} can be easily read using Python's Pandas library. An example code can be found in the scripts used for creating the plots in this paper and also the feature extraction and regression. In order to use the raw audio recordings which are stored as wav files, we recommend using the Python's Librosa library. An example of this procedure is found in the scripts related to feature extraction step available at our Github repository (\url{https://github.com/MuSAELab/UrBAN}). 

This dataset provides different labels, as detailed in Table~\ref{tab:inspections-summary}. This can enable the development of different supervised learning tasks. Moreover, given the advances seen with unsupervised learning, specifically self-supervised learning (SSL) of audio signals \cite{niizumi2021byol-a}, the dataset can open door for numerous other applications. The work in \cite{arxiv_1, 10424387}, for example, used SSL techniques to detect beekeeper speech in the beehive audio recordings to improve hive monitoring performance. It is hoped that the UrBAN dataset will enable new applications that can improve the work of the beekeepers and the lives of the honey bees. 

\section*{Code availability}
The code used for creating the plots, audio enhancement, feature extraction, and frames of bees prediction are all categorized and available at our Github repository (\url{https://github.com/MuSAELab/UrBAN}). 

\bibliography{references}

\section*{Acknowledgements} 

The authors acknowledge funding from NSERC via their Alliance program (ALLRP 548872-19), as well as Nectar Technologies Inc and the Centre de recherche en sciences animales de Deschambault for the support with data collection. 

\section*{Author contributions statement}

M.A. contributed to the initial drafting of the manuscript and conducted the experiments in the technical validation. Y.Z. and H.G. contributed to the feedback to experimental results. N.C. contributed to sensor data acquisition and verified data records. S.M. contributed in editing and review of the manuscript. P.G. and T.F. contributed in conceptualizing the study and study design, and supervision of the study. All authors contributed to the critical revision of the manuscript. All authors had full access to all the data in the study and took responsibility for the decision to submit this draft for publication.

\section*{Competing interests}
The authors declare no competing interests.

\end{document}